\documentclass{JHEP3} % 10pt is ignored!

\usepackage{epsfig}

\title{The Atick-Witten free energy, closed  tachyon \\
       condensation and deformed Poincar\'e symmetry}
\author{Michele Maggiore\\
D\'epartement de Physique Th\'eorique, Universit\'e de Gen\`eve,\\
24 quai Ernest-Ansermet, CH-1211 Gen\`eve 4\\
E-mail: \email{michele.maggiore@physics.unige.ch}}
\abstract{The dependence of the free energy of string theory on the
temperature at $T\gg T_{\rm Hag}$ was found long ago by Atick and
Witten and is  $F(T)\sim\L T^2$,  
where $\L$ diverges because of a tachyonic instability.
We  show that this result can be
understood assuming that, above the Hagedorn  transition,
Poincar\'e symmetry is deformed into a
quantum algebra. Physically this quantum algebra
describes a non-commutative spatial geometry
and a discrete euclidean time.  We then show that in string theory
this deformed Poincar\'e symmetry
indeed emerges above the Hagedorn temperature from the condensation of
vortices on the world-sheet.
This result indicates that the endpoint of the condensation
of closed string tachyons with non-zero winding
is an infinite stack of spacelike branes with a given
non-commutative world-volume geometry.
On a more technical side, we also point out that $T$-duality
along a circle with antiperiodic boundary conditions for spacetime
fermions is broken by world-sheet
vortices, and the would-be T-dual variable becomes non-compact.
}
\keywords{Strings at finite temperature, Tachyon condensation,  Quantum groups}
\preprint{UGVA-DPT 2002/02-1098}
\renewcommand\({\left(}
\renewcommand\){\right)}
\renewcommand\[{\left[}
\renewcommand\]{\right]}

\newcommand{\ra}{\rightarrow}

\def\lsim{\raise 0.4ex\hbox{$<$}\kern -0.8em\lower 0.62
ex\hbox{$\sim$}}

\def\gsim{\raise 0.4ex\hbox{$>$}\kern -0.7em\lower 0.62
ex\hbox{$\sim$}}

\newcommand\eq[1]{eq.~(\ref{#1})}
\newcommand\eqs[2]{eqs.~(\ref{#1}) and (\ref{#2})}

\newcommand\pa{\partial}

\newcommand\ee{\end{equation}}
\newcommand\be{\begin{equation}}
\def\bea{\begin{eqnarray}}
\def\eea{\end{eqnarray}}
\newcommand\ees{\end{eqnarray}}
\newcommand\bees{\begin{eqnarray}}

\newcommand{\thag}{T_{\rm Hag}}

\def\dslash{\not{\hbox{\kern-2pt $\partial$}}}
\def\Dslash{\not{\hbox{\kern-4pt $D$}}}
\def\pslash{\not{\hbox{\kern-2.3pt $p$}}}
\def\ep{\epsilon}
\def\eps{\epsilon}
\def\f{\phi}

\def\O{\Omega}
\def\D{\Delta}

\def\a{\alpha}
\def\b{\beta}

\def\th{\theta}
\def\s{\sigma}
\def\g{\gamma}

\def\L{\Lambda}
\def\d{\delta}

\newcommand\fverb{\setbox\pippobox=\hbox\bgroup\verb}
\newcommand\fverbdo{\egroup\medskip\noindent%
			\fbox{\unhbox\pippobox}\ }
\newcommand\fverbit{\egroup\item[\fbox{\unhbox\pippobox}]}
\newbox\pippobox
\begin{document}

\section{Introduction}

One of the long-standing problems of string theory is to understand
what happens  above the Hagedorn temperature $\thag$. The existence of
many analogies with the deconfinement transition in QCD suggests that 
at $T>\thag$ it could be possible to discover
a more fundamental level of description
and   more fundamental degrees of freedom. 

One of the best hints for understanding the physics above $\thag$ comes
from the Atick-Witten computation of the closed strings
free energy~\cite{AW}.
In the limit $T\gg\thag$ the result, both for the bosonic string and
for superstrings, is
\be\label{T2}
F(T)\ra V\L T^2
 \, ,\hspace{15mm} (T\ra\infty )
\ee
where $V$ is the spatial volume and $\L$ a divergent cosmological constant.
This should be
compared with the behavior of the free energy
 of field theory in $D$ space-time dimensions,
$F(T)\sim VT^D$, which would have rather suggested $F(T)\sim VT^{26}$
for the bosonic string and $F(T)\sim VT^{10}$ for superstrings.
Eq.~(\ref{T2}) seems therefore to
indicate a vast reduction in the number of degrees of
freedom. 

The divergence of $\L$ is related to the fact that at $\thag$ a
tachyonic instability develops. Understanding the physics behind
\eq{T2} means therefore understanding what is the endpoint of the
condensation of this closed string tachyon.
If one is able to identify the correct
vacuum, it becomes possible to compute  around it, and to find a
finite value for $\L$.

In this paper we will  show that \eq{T2}  expresses the fact that above
$\thag$ there is a phase transition and Poincar\'e symmetry is
deformed, i.e. it becomes a quantum algebra with a deformation
parameter $a$ with dimensions of length. Physically, this quantum
algebra describes a system in which 
the fictitious euclidean time used to study finite temperature
becomes discrete, and $a$ is the corresponding lattice spacing.

The physics of this 
deformation of Poincar\'e symmetry will be discussed in
sect.~\ref{sect.aP}. 
The quantization of a system described by this deformed algebra
is particularly
interesting, and we will find that it
automatically implies a non-commutativity between the spatial
coordinates, as well as
a generalized uncertainty principle between coordinates and momenta.
In fact, we will just reobtain the generalized
uncertainty principle that we found in ref.~\cite{MM1}.
Following a  recent suggestion \cite{KR}, we will then see 
that in a system with deformed Poincar\'e symmetry, in the limit where
$T$ is much larger than any scale and in particular than the deformation scale,
 $aT\gg 1$, the free energy comes out automatically of the form
(\ref{T2}), in any number of dimensions;
$\L$ is finite, and is related to the deformation parameter $a$,
$\L\sim (1/a)^{D-2}$.

In sect.~\ref{sect4}
we apply these ideas to the Hagedorn transition  in string theory.
We will first  recall in sect.~\ref{sect2} some
basic facts about the Atick-Witten result, emphasizing some aspects
(as the lack of thermal duality for superstrings) that will play an
important role in our analysis.
Our main goal in sects.~\ref{sect.ws}--\ref{sect.tt}
will  be to show that the deformation of Poincar\'e
symmetry discussed in sect.~\ref{sect.aP} indeed
emerges in string theory
from the world-sheet dynamics. A crucial role will be played
by world-sheet vortices. It is well known~\cite{GK1,GK2} 
that the discretizations
of the action of a compact
string coordinate
\be\label{1S}
S=\frac{1}{4\pi\a'}\int d^2\xi\,\, (\pa_{\a} X )^2\, ,\hspace{10mm}
X\sim X+2\pi R
\ee
fall into two different universality classes, depending on whether
they admit or not vortices. The issue is irrelevant at large $R$,
since  vortices are anyway dynamically suppressed, but becomes
crucial below a critical value  $R_c$, where world-sheet vortices, if
included, 
dominate the partition function.
We must therefore ask what is the correct definition, in
the application to string theory,  of the 
formal continuum action (\ref{1S}).
In sect.~\ref{sect.ws} we will show that in superstring theory the
correct prescription is that  vortices should not be included
when, compactifying
along $X$, we impose periodic
boundary conditions both on spacetime bosons and fermions, 
while we must include them if we instead impose antiperiodic boundary
conditions on fermions, i.e if we compactify on $S^1/(-1)^{\bf F}$ with
${\bf F}$ the spacetime fermion number. The difference 
in the prescription between
$S^1$ and  $S^1/(-1)^{\bf F}$ has its roots in the fact that
in these two cases there are also two
different prescriptions for the GSO projection~\cite{AW,Rohm}. 
In particular, vortices must be retained when we compactify $X^0$
to study finite temperature, since in this case we 
impose the $(-1)^{\bf F}$ twist. 

Retaining the vortices in the compactification of
$X^0$ (or of any variable compactified with a  $(-1)^{\bf F}$ twist)
has crucial effects on the world-sheet dynamics
below a critical radius $R_c$. At $R=R_c$ there is in fact 
on the world-sheet a Kosterlitz-Thouless (KT) phase
transition driven by vortex condensation.
When $X=X^0$ is the euclidean
time direction, $1/(2\pi R_c)$ coincides with the Hagedorn
temperature~\cite{Sat,Kog,OBT,AW}, and therefore this phase transition
corresponds, on the world-sheet, to the Hagedorn phase transition
 in spacetime.

Using standard results from the renormalization group analysis of the
KT phase transition, we will see that below $R_c$ vortices have two crucial
effects. 
First, T-duality is broken, and 
the variable $\tilde{X}^0=X_L^0-X_R^0$ that usually 
describes the T-dual physics and lives on
a circle with dual radius $R'=\a'/R$ becomes uncompactified,
and  lives now on the whole real line. Second, vortices generate an
effective potential for $\tilde{X}^0$ 
\be
V(\tilde{X}^0)= -\mu\cos \(\frac{R}{\a'}\, \tilde{X}^0\)\, ,
\ee
with $\mu$ going to infinity 
under renormalization group transformations
on the world-sheet. This
potential breaks the translation invariance of  $\tilde{X}^0$  and
localizes it on the minima of the cosine, and  therefore gives
rise to a lattice spacing  
$a = 2\pi\a '/R$. This is in full agreement with results
already obtained some time ago
by Gross and Klebanov~\cite{GK1,GK2} using matrix model techniques.
We will confirm this result also from a $\s$-model analysis of 
tachyon condensation in sect.~\ref{sect.tt}.

In turn this will allow us to put forward a definite proposal for the
condensation of closed string tachyons; we will claim that the
endpoint of the condensation of  closed string tachyons with 
non-zero winding is a stack of spacelike branes, with a
non-commutative world-volume geometry fixed by the deformation of the
Poincar\'e algebra. It would be very interesting to understand
whether these branes have a CFT description in terms of 
strings with Dirichlet boundary conditions in the 
temporal direction, along the lines discussed recently 
in refs.~\cite{Hull,GS2}.
We will also compare this picture with
recent results on the condensation of  type~0 
tachyons~\cite{CG,GS,Adams,David}, and we will
find elements which support our proposal.

\section{The physics of deformed Poincar\'e symmetry}\label{sect.aP}

\subsection{Deformed algebras and discretized spacetime} \label{sectdef}

Let us begin by considering, for simplicity, a 1+1 dimensional system
with continuous time and discrete space, with lattice spacing $a$. 
In this setting a Klein-Gordon equation reads
\be\label{KG1}
\( -\pa_t^2 +\D_x^2 -m^2 \)\f =0
\ee
with  $\D_x\f (t,x) =[\f (t, x+a)-\f (t,x-a)]/2a$.
The dispersion relation that follows is
\be\label{disp2}
E^2=\frac{\sin^2 ap}{a^2} +m^2\, . 
\ee
The physics of \eq{disp2}
is clear. Momentum is periodically identified, and  the lattice can
substain  travelling waves only up to a maximum energy 
$E_{\rm max}=(a^{-2}+m^2)^{1/2}$. If we  replace $c=1$ with
 a speed $v<1$, this
equation just describes phonons in $1+1$ dimensions.
From a Lie algebra point of view, the symmetry of the system is
described by the generator $H$ of continuous time translations and the
generator $P$ of discrete spatial translations,
satisfying the Lie algebra $[H,P]=0$, supplemented by the
identification $P\sim P+2\pi/a$. The symmetry under
rotations in the $(t,x)$ plane, i.e. boosts, is broken, and no
generator is associated to it.

There is however an alternative description of the symmetry of this
system, based on a deformed algebra~\cite{Cel}. One introduces
also the boost generator $K$, and considers the
algebra
\be\label{PH}
[P,H]=0\, ,\hspace{10mm} [K,P]=iH\, ,\hspace{10mm}
[K,H]=i\,\frac{\sin (2aP)}{2a}\, .
\ee
In the limit $a\ra 0$ this reduces to the standard
Poincar\'e algebra of a 1+1
continuous relativistic system. 
The structure (\ref{PH}) is however
well defined even at finite $a$, because it is easy to see
that the commutators indeed obey the Jacobi identities\footnote{In
fact, we can equate the commutator $[K,H]$ to an arbitrary function
of $P$, and still the Jacobi identities are trivially satisfied. In
this way we can obtain an algebra corresponding to an arbitrary
discretization of the spatial derivative.}. 
Eq.~(\ref{PH}) is an example of a {\em quantum
algebra}, or {\em deformed algebra}; $a$ is the deformation parameter.

The physical relevance of this 
construction emerges from the observation
that this quantum algebra has a quadratic Casimir $C_2$ given by
\be
C_2 =H^2-\frac{\sin^2 (aP)}{a^2}\, ,
\ee
as well as the realization, in position
space
\be
P_{\mu}=-i\pa_{\mu}\, ,\hspace{10mm}
K=ix\, \pa_t -t\, \frac{\sin (-2ia\pa_x)}{2a}\, ,
\ee
where $P^{\mu}=(H,P)$ and we use $\eta_{\mu\nu}=(-,+)$. 
Therefore the discrete KG equation (\ref{KG1}), or equivalently the dispersion
relation (\ref{disp2}),  is simply the condition
$C_2=m^2$, and in this sense this deformed Poincar\'e algebra can be
considered as the symmetry of a relativistic system living in discrete
one-dimensional space and
continuous time.

Comparing the Lie algebra and the deformed algebra descriptions of
this system
we see that
in the Lie algebra approach, when $a\neq 0$, there are only two
generators, $H$ and $P$; $a=0$ is a point of enhanced
symmetry, where a new generator $K$ suddenly
pops out. In the deformed algebra
description instead we always have the 
three generators $H,P,K$ even for finite
$a$, so in a sense we always have the information about the
existence of a symmetry group with three generators,
but we pay this with a non-linear structure. The point $a=0$ is the
point where the algebraic structure linearizes. 

At this classical level, however, the Lie algebra  and the deformed
algebra decriptions of this system contains basically the same amount of
informations.\footnote{Both when we consider
a single particle system and composite systems. In the case of 
composite systems there is some confusion in the literature, and we
clarify the issue 
in the appendix~\ref{sectcomp}.}  The dynamics of the classical
system is completely specified by its equation of motion.
For the deformed algebra, this comes out from the quadratic
Casimir, while from the Lie algebra point of view \eq{KG1} reflects the
covariance under continuous time translation and discrete spatial
translations.

\vspace{5mm}

Our real interest, however, is in systems with discrete {\em time} and
continuous space, and in this case
we will find in sect.~\ref{sect3} that, after quantization,
the description based on  the deformed algebra  leads to very substantial 
differences from the description based on the Lie algebra.
Consider therefore a system, again for the moment 1+1 dimensional and
with minkoskian signature, in which  time is discrete and space is
continuous. The KG equation reads $(-\D_t^2+\pa_x^2-m^2)\f =0$ and the
dispersion relation is
\be\label{disp0}
\frac{\sin^2 aE}{a^2}= p^2+m^2\, .
\ee
Even if our starting point, a KG with a finite time difference, seemed
reasonably simple, the physics of this dispersion relation is quite
peculiar, and  there is  a maximum allowed momentum, and
even a maximum mass, $p^2+m^2\leq 1/a^2$. 
It is easy to find a quantum algebra description of this system,
simply exchanging the role of $H$ and $P$ in \eq{PH}, 
\be\label{HP}
[P,H]=0\, ,\hspace{10mm} [K,H]=iP\, ,\hspace{10mm}
[K,P]=i\,\frac{\sin (2aH)}{2a}\, .
\ee
and again  the KG equation is reproduced by the Casimir, which now is
\be
C_2 =\frac{\sin^2 (aH)}{a^2}-P^2\, .
\ee
A third deformation of the Poincar\'e algebra, which will turn out to
be the most relevant for our purposes, is obtained discretizing {\em
euclidean time} and then rotating back into Minkowski.
In this
case we start from  an euclidean  KG equation
$(\D_t^2+\pa_x^2-m^2)\f =0$, leading to a dispersion relation
$-(\sin^2 aE)/a^2= p^2+m^2$.
When  we rotate  back  into Minkowski space, $E\ra -iE$,
the dispersion relation becomes
\be\label{disp}
\frac{\sinh^2 aE}{a^2}= p^2+m^2\, .
\ee
Of course both \eq{disp} and \eq{disp0} reduce to the standard
Minkoskian dispersion relation $E^2=p^2+m^2$ in the
limit $a\ra 0$. However, at finite $a$ they are different and
they describe very different physics.
For instance, in \eq{disp0}
there is maximum  momentum, which is not the case 
for  \eq{disp}. Therefore physically
a system with a discrete minkoskian time, \eq{disp0}, has nothing to
do with the system obtained discretizing first time in euclidean space
and then  rotating back into Minkoswki space.
However formally \eqs{disp0}{disp} are  related by $a\ra ia$.
Substituting $a\ra ia$ into \eq{HP} we therefore
find a deformation of the Poincar\'e algebra whose
Casimir reproduces the dispersion relation (\ref{disp}),
\be\label{euP}
[P,H]=0\, ,\hspace{10mm} [K,H]=iP\, ,\hspace{10mm}
[K,P]=i\,\frac{\sinh (2aH)}{2a}\, .
\ee
Eqs.~(\ref{HP}) and (\ref{euP}) are special cases of deformations of
the Poincar\'e algebra that can be written in any number of
dimensions~\cite{LRNT,LNR,LR}. The deformation relevant for a 
system with $d$ spatial dimensions, in Minkowski space, with discrete
Minkowski time, (i.e. the generalization of (\ref{HP}) ) is
as follows: all commutators involving the angular momentum $J_{ij}$
are the same as in the undeformed
Poincar\'e algebra. Hence, the group of spatial
rotations is not deformed. Similarly for spacetime
translations still holds 
$[P_{\mu},P_{\nu}]=0$. The commutators involving the boosts
$K_i=J_{i0}$  are instead
\be\label{lnr1}
[K_i,H]=iP_i\, ,\hspace{5mm} [K_i,P_j]=i\d_{ij}
\frac{\sin (2aH)}{2a}\, ,
\ee
\be\label{lnr2}
[K_i,K_j]=-iJ_{ij}\cos (2aH)-
ia^2P^k(P_iJ_{jk}+P_jJ_{ki}+P_kJ_{ij})\, .
\ee
The quadratic Casimir is
\be
C_2 =\frac{\sin^2 (aH)}{a^2}-{\bf P}^2\,  .
\ee
The deformation relevant instead for a
system with $d$ spatial dimensions, obtained discretizing first
euclidean
time and then rotating back to Minkowski is obtained with $a\ra ia$
and is therefore~\cite{LNR,LR}
\be\label{elnr1}
[K_i,H]=iP_i\, ,\hspace{5mm} [K_i,P_j]=i\d_{ij}
\frac{\sinh (2aH)}{2a}\, ,
\ee
\be\label{elnr2}
[K_i,K_j]=-iJ_{ij}\cosh (2aH)+
ia^2P^k(P_iJ_{jk}+P_jJ_{ki}+P_kJ_{ij})\, .
\ee
with all other commutators undeformed, and with
\be
C_2 =\frac{\sinh^2 (aH)}{a^2}-{\bf P}^2\,  .
\ee

\subsection{Quantization: non-commutative geometry
and  generalized uncertainty principle}\label{sect3}

The real difference between the Lie algebra and the deformed algebra
description of a system with discrete time appears when we quantize the system.
Compare in fact
what happens in the two cases (\ref{disp2}) (discrete space) and
 (\ref{disp0}) (discrete Minkowski time) when we
quantize a particle imposing 
$[x_i,p_j]=i\d_{ij}$. In momentum space the operator $x_i$ is 
represented as $i\frac{\pa}{\pa p_i}$, and the velocity of the
particle in the Heisenberg representation is given by
\be
\dot{x}_i=i[H,x_i]=\frac{\pa E}{\pa p_i}\, .
\ee
In the  familiar case of a phonon this  of course gives
the standard expression for the group velocity: using \eq{disp2} 
 (and setting for simplicity $m=0$), one gets
\be
\frac{\pa E}{\pa p} = \cos (ap)\, ,
\ee
i.e. the standard group velocity of a massless particle 
on a regular spatial lattice.
Instead, for a lattice in Minkowski time, using \eq{disp0} we get
\be\label{3.5}
\frac{\pa E}{\pa p_i} = \(\frac{p_i}{\sqrt{p^2+m^2}}\)\, \frac{1}{\cos
(aE)}\, .
\ee
The term in parenthesis is just the standard expression for the
velocity in terms of momentum. However, the cosine at the denominator 
makes no sense, and if we would take \eq{3.5} as an expression for the
velocity we would find $v>1$, and even $v\ra\infty$, when $aE$
approaches $\pi/2$, i.e. near the maximum momentum.

Clearly something has gone wrong, and we cannot quantize in this way
a particle on a space with discrete Minkowski time.
The simplest attitude would be to say
that, when time is discrete, time derivatives make no sense and should
be replaced by  finite differences, so  we must give up equations like
$v_i=i[H,x_i]$. This is the approach that
would be taken within the Lie algebra description of the symmetries of
the system, and it is completely analogous to the fact that 
in the Lie algebra approach the boost
generator $K$ only appears at $a= 0$.

The description of the symmetries in terms of a deformed algebra
suggests however a different route. 
Namely, we do not give up the possibility of having boosts, and
therefore velocities, even at finite $a$, but we accept to pay this
with the introduction of a non-linear structure.

In particular, we can retain the equation $v_i=i[H,x_i]$, but we
modify the definition of the
position operator requiring that the relation between 
velocity and momentum is not deformed,\footnote{This 
choice for $v_i$ is physically
very reasonable, and 
the deeper reason for it is explained in app.~\ref{sectcomp}.} 
$v_i=p_i/\sqrt{p^2+m^2}$.
This is easily done defining, in momentum space,
\be\label{3.6}
x_i=i \cos(aE)\frac{\pa}{\pa p_i}=
i  \sqrt{1-a^2({\bf p}^2+m^2)} \,\,\frac{\pa}{\pa p_i}\, ,
\ee
where in the second equality we used the dispersion relation.
By construction we now have
\be
v_i=i[H,x_i]=\cos (aE) \frac{\pa E}{\pa p_i}
=\frac{p_i}{\sqrt{p^2+m^2}}\, ,
\ee
and $x_i$ obviously has the correct limit for $a\ra 0$. However, 
using \eq{3.6}, we can compute explicitly the $[x_i,x_j]$ and 
$[x_i,p_j]$ commutators, and we find
\bees
\[ x_i, x_j\] 
& = & ia^2J_{ij}\, ,\label{xxm}\\
\[ x_i, p_j\] & = & i \,
\delta_{ij}\,\, \sqrt{1-a^2({\bf p}^2+m^2)}\, ,\label{xpm}
\ees
where we have defined 
\be\label{J}
J_{ij}=-i\( p_i\frac{\pa}{\pa p_j}-p_j\frac{\pa}{\pa p_i}\)\, .
\ee
Note that $J_{ij}$ satisfies the usual commutation relations of
angular momentum, so that the group of spatial rotations is not
deformed.

This result is quite surprising; first of all, space is now
non-commutative, and \eq{xxm} is just of the type
first discussed, long ago, by Snyder~\cite{Sny}.\footnote{More
precisely, in ref.~\cite{Sny} there was also a non-commutativity
between time and spatial coordinates, whose commutator closed on the
boosts, which is not the case for us.} Second, the
uncertainty principle is modified. Amazingly, at the maximum allowed
value of the momentum the $[x,p]$ commutator vanishes.

These commutation relations are quite interesting by themselves, but they are
not yet the setting that will be used to reproduce the Atick-Witten 
free energy. A partition function 
at finite temperature $T$ is formally equivalent to the path integral
for a system with {\em imaginary} ``time'' ranging from zero to $1/T$, and it
is this fictitious euclidean time that we want to discretize.
As we discussed in sect.~\ref{sectdef}, this is
obtained from the case with discrete Minkowski time simply
with the replacement
$a\ra ia$. Then the dispersion relation is \eq{disp}, 
the position operator becomes\footnote{A more precise analysis uses the
deformation of the Newton-Wigner position operator, which is hermitean
with respect to the scalar product invariant under deformed Poincar\'e
symmetry. This analysis is explained in detail in ref.~\cite{MM3}, but the
final result for the $[x,x]$ and $[x,p]$ commutators is the same.}
\be
x_i=i\cosh (aE)\frac{\pa}{\pa p_i}
\ee
 and
the commutation relations become
\bees
[x_i,x_j] &= &-ia^2J_{ij}\, ,\label{xxe}\\
\[ x_i,p_j\] & = & i\,\delta_{ij}\,\, \sqrt{1+a^2({\bf p}^2+m^2)}
\, .\label{xpe}
\ees
Eq.~(\ref{xpe}) shows that in this case at large energies the
volume of the cells of the phase space increases. Expanding at first
order in $a^2$ one finds~\cite{MM1} that
\eq{xpe} implies a generalized uncertainty principle of
the form
\be\label{3.14}
\Delta x\geq \frac{1}{\Delta p}+ a^2\Delta p
\ee
and therefore a minimum 
observable length of order $a$. A generalized uncertainty
principle of this form has been obtained in
string theory, studying planckian scattering in
the eikonal limit \cite{Ven,ACV,GM,Ken}, and also
in quantum gravity from gedanken black-hole experiments~\cite{MM2}.
We will see that \eq{3.14}, or better the full expression
(\ref{xpe}), is also relevant in the high temperature regime of string
theory.

The commutators (\ref{xxm}, \ref{xpm}) and  (\ref{xxe}, \ref{xpe})
where obtained some time ago~\cite{MM1} using the following argument,
that illustrates their uniqueness.
Suppose that we look for the most general deformation of the Heisenberg
algebra, $[x_i,x_j]=0$, $[x_i,p_j]=i\delta_{ij}$, in $d$ spatial
dimensions, which depends on a deformation parameter $a$ with dimensions
of length, and 
such that for $a\ra 0$ we recover the standard Heisenberg algebra.
Impose further  the conditions that the group of spatial rotations is
not deformed, so that the $J_{ij}$'s still close the standard algebra of
$SO(d)$, and also that the group of spatial translations is not deformed,
so that $[p_i,p_j]=0$. We then look for the most general deformed
algebra that can be
constructed using only $x_i,p_i$ and $J_{ij}$.
The commutator $[x_i,x_j]$ can only be proportional to $J_{ij}$,
since it is the only available antisymmetric tensor with the same
transformation properties under rotation (and  parity). The
proportionality factor $a^2$ is dictated by dimensional considerations
and a factor of $i$ by hermiticity, so the most general
form is
\be
[x_i,x_j]= ia^2g({\bf p}^2)\, J_{ij}\, ,
\ee
with $g$ an arbitrary real function. Invariance under rotations and
translations requires that $g$ can depend only on ${\bf p}^2$
(in particular, a dependence on ${\bf x}^2$ or ${\bf x\cdot p}$ is
forbidden by translation invariance).
Similarly,  invariance under spatial rotations
and translations requires a general form
\be\label{xpf}
[x_i ,p_j]=i f({\bf p}^2)\, \delta_{ij}\, .
\ee
The remarkable fact is that  the Jacobi identities fix uniquely
the functions $f({\bf p}^2), g({\bf p}^2)$. Consider first
the identity with three $x$'s,
\bees
0 &=& [x_i,[x_j,x_k]]+{\rm cyclic}=
 [x_i, g({\bf p}^2) J_{jk}]+[x_j, g({\bf p}^2)  J_{ki}]+
[x_k, g({\bf p}^2) J_{ij}] =\nonumber\\
&=& g({\bf p}^2) \( [x_i,  J_{jk}]+[x_j,   J_{ki}]+
[x_k,  J_{ij}] \) +
\( [x_i, g] J_{jk}+[x_j, g]  J_{ki}+
[x_k, g] J_{ij} \)\, .
\ees
The first parenthesis in the last line vanishes automatically, using
the fact that the rotation
group is undeformed, so that it still holds $[J_{ij},V_k]=i(\d_{ik}V_j-
\d_{jk}V_i)$, for any vector $V_i$.
To compute $[x_i,g]$ observe that \eq{xpf} implies that in momentum
space $x_i$ can be represented as 
\be\label{xfp}
x_i=i f({\bf p}^2) \frac{\pa}{\pa p_i}\, .
\ee
Then we get 
\be
0=2if\frac{\pa g}{\pa {\bf p}^2}\( p_iJ_{jk}+p_jJ_{ki}+p_kJ_{ij}\)
\ee
The term $( p_iJ_{jk}+p_jJ_{ki}+p_kJ_{ij})$ vanishes automatically if
$J_{ij}$ is the {\em orbital} angular momentum (\ref{J}),
i.e. for spin zero particles. However, it
is non zero on a generic representation of the rotation group. 
Since the Jacobi identities must hold independently of the
representation,  we must have either $f=0$ or 
$g=$~const. The choice
$f=0$ of course does not reproduce the standard commutators in the
limit $a\ra 0$, so we conclude that $g$ is a constant. With a
redefinition of $a$, we can set it to $\pm 1$. We therefore conclude
that the most general result for the $[x,x] $ commutator is
\be
[x_i,x_j]= \pm ia^2 J_{ij}\, .
\ee
The other Jacobi identities now fix the function $f$.
The Jacobi identity with three $p$'s and that with one $x$ and two $p$'s
are trivially satisfied as a consequence of $[p_i,p_j]=0$. 
The last Jacobi identity is
\be
0=[x_i,[x_j,p_k]] +[x_j,[p_k,x_i]] + [p_k,[x_i,x_j]]
 =\d_{jk}[x_i,f({\bf p}^2)] -
 \d_{ik}[x_j,f({\bf p}^2)]
\pm a^2 [p_k,J_{ij}]\, .\label{last}
\ee
Using again \eq{xfp}, \eq{last} becomes
\be
2f\frac{\pa f}{\pa {\bf p}^2}\, (\d_{jk} p_i -\d_{ik}p_j)
=\mp a^2  (\d_{jk} p_i -\d_{ik}p_j)\, ,
\ee
or
\be
\frac{\pa f^2}{\pa {\bf p}^2}=\mp a^2\, .
\ee
The solution (imposing further that $f$ is actually a function of the
combination ${\bf p^2}+m^2$ rather than only of ${\bf p}^2$, and that
$f=1$ when $a=0$), is
\be
f({\bf p}^2) = \[ 1\mp a^2 ({\bf p}^2+m^2)  \]^{1/2}\, ,
\ee
which reproduces \eqs{xpm}{xpe}.

In general, it is not obvious that a deformation of a given Lie algebra
exists, since the Jacobi identities provide very stringent requirements
on non-linear structures. Here we see, first of all, that it is
possible to deform the Heisenberg algebra with a dimensionful parameter
and, second, that this deformation is {\em unique}, modulo the
replacement
$a\leftrightarrow ia$, and within the rather general assumptions
that we have discussed.
Furthermore, the functions $f({\bf p}^2)$ 
and $g({\bf p}^2)$
are the same in any number of spatial dimensions.

Finally we observe that, with a non-local redefinition of coordinates,
$x_i\ra x_i/f({\bf p}^2)$, we can always reduce the $[x,p]$ commutator
to the standard form  $[x_i,p_j]=i\delta_{ij}$.
However, we would pay this at the level of the
action, which would become non-local if we started from a local
expression, 
or equivalently at the level of the dispersion relation. Thus, 
once we say that our starting point is, e.g., a Klein-Gordon equation 
$(-\D_t^2+\pa_i^2-m^2)\f =0$, 
the physical definition of coordinates   and momenta
has been fixed
and we find that there is a non-trivial
deformation of the Heisenberg algebra, and this deformation is unique,
with the assumptions made.

\subsection{Statistical mechanics:
 $F(T)= \L VT^2$}\label{sect3.1}

A deformed quantization procedure implies  a deformation of the
standard rules of statistical mechanics. The following important 
observation has been made recently 
by Kalyana Rama \cite{KR}. Consider a system in $d$ spatial
dimensions ($d=D-1$), described by the deformed Poincar\'e algebra with
dispersion relation (\ref{disp}) and therefore
quantized according to the
deformed commutation relations (\ref{xxe}, \ref{xpe}). 
The volume of the cells of the phase space is  not anymore 
$(2\pi \hbar )^d$, but rather $(2\pi \hbar f(E) )^d$, where 
\be
f(E)\equiv  \sqrt{1+a^2({\bf p}^2+m^2)} =\cosh (aE)\, .
\ee
As a consequence, all  averages over the phase space
should be performed using this
new measure,
\be
\int \frac{d^dqd^dp}{(2\pi\hbar )^d}\,\( \cdots \)
\ra
\int \frac{d^dqd^dp}{[2\pi f(E)\hbar ]^d}\,\( \cdots \)\, .
\ee
Let us first use for illustration the 
Maxwell-Boltzmann statistics; the free energy
 at temperature $T$ 
of such a system is   given by  
\be\label{Z1}
-\frac{F(T)}{T}=\int \frac{d^dqd^dp}{[2\pi f(E) ]^d}\,\, 
e^{- E/T} 
= V\frac{\O_{d}}{(2\pi )^d}
\int \frac{ p^{d-1}dp}{\cosh^d (aE)}\,\, e^{- E/T}\, ,
\ee
where we have set again $\hbar =1$, and $\O_{d}$ is the solid angle.

In string theory
at $T\ll \thag$ the  free energy is dominated by the massless
string modes and reproduces the standard field theoretic behavior 
$F(T)\sim VT^D$, plus exponentially small corrections from the massive
modes. As $T$ approaches $\thag$, the cumulative effect of the massive
modes becomes important, because of the exponentially raising density
of states and, above the transition, it takes over.
The result $F(T)\sim VT^2$ therefore
can be seen as a consequence of the cumulative effects of all
string massive modes. To understand physically this result means to be
able to explain it in terms of the effect of a few new
{\em light} modes,
which would then be the more appropriate degrees of freedom for
describing the new phase.

Since we aim at explaining the result for the free energy in
terms of light modes, we can neglect the
mass term and write the dispersion relation simply as 
\be\label{disp4}
p=\frac{1}{a}\sinh aE\, ,
\ee
and $dp=\cosh (aE)dE$. Then \eq{Z1} becomes
\bees
-\frac{F(T)}{T}
&=&V\frac{\O_{d}}{(2\pi )^d a^{d-1}}
\int_0^{\infty} dE\, (\tanh aE)^{d-1}\, e^{- E/T}=\nonumber\\
&=& VT\frac{\O_{d}}{(2\pi )^d a^{d-1}}
\int_0^{\infty} dx\, [\tanh (aTx)]^{d-1}\, e^{- x}\, .
\ees
In the limit $aT\ra\infty$ at fixed $x$ we have $\tanh (aTx)\ra 1$ and
since the integrand is regular near $x=0$ we can take the limit inside
the integral,
\be
\lim_{aT\ra\infty}\int_0^{\infty} dx\, [\tanh (aTx)]^{d-1} e^{- x}=
\int_0^{\infty} dx\,  e^{- x}=1\, .
\ee
Correspondingly, one finds at large $T$~\cite{KR},
\be\label{3.30}
F(T)\simeq - \(\frac{\O_{d}}{(2\pi )^d a^{d-1}}\)\, VT^2\, .
\ee
Note that the $T^2$ dependence holds in
any number of spatial dimensions since,  apart from the overall
constant, $d$  appears only in
$ [\tanh (aTx)]^{d-1}$, which saturates to 1.
The physical mechanism behind this result is
that the growth of the volume of the cells of the phase space,
combined with the modified dispersion relation, compensates exactly the
phase space factor $p^{d-1}dp$.

The behavior  $F(T)\sim VT^2/a^{d-1}$ 
can be obtained also using Bose-Einstein or Fermi-Dirac
statistics. For Bose-Einstein the result follows from
\be\label{BE}
\lim_{aT\ra \infty}\int_0^{\infty} dx\, [\tanh (aTx)]^{d-1} 
(-1) \log (1-e^{- x}) =-\int_0^{\infty} dx\,\log (1-e^{-
x})=\frac{\pi^2}{6}
\ee
and for Fermi-Dirac 
\be\label{FD}
\lim_{aT\ra \infty}\int_0^{\infty} dx\, [\tanh (aTx)]^{d-1} 
 \log (1+e^{- x}) =\int_0^{\infty} dx\,\log (1+e^{-
x})=\frac{\pi^2}{12}\, .
\ee
Therefore the result for
the  free energy in the large $T$ limit is
\be\label{FT2}
F(T)\ra \L \, VT^2\, ,\hspace{15mm} (aT\ra\infty )
\ee
with
\be\label{L1a}
\L =-\( (2n_B+n_F)\frac{\pi^2}{12}\,\frac{\O_d}{(2\pi)^{d} }\)\,\, 
\frac{1}{a^{d-1}}\, ,
\ee
where $n_B$ is the number of light boson species and $n_F$ the number
of light fermion species that constitute the relevant degrees of
freedom in the  phase with deformed Poincar\'e symmetry.

If we remove the deformation parameter, 
i.e. $a\ra 0$, while keeping $T$ fixed, we cannot use \eqs{BE}{FD},
since they have been obtained in the
limit $aT\ra\infty$. Instead, 
if $a\ra 0$ at fixed $T$
 we obviously reobtain the
undeformed result $F(T)\sim VT^D$. However, if we first  take the large
$T$ limit at fixed $a$ and then we take the limit $a\ra 0$ in
\eq{L1a}, then we see that  $\L$
diverges. As we will recall in sect.~\ref{sect2}, in string theory,
taking first the large $T$ limit,  one
finds a result just of the form (\ref{FT2}), 
with $\L$ divergent, because of
a tachyonic instability.
This divergence must disappear if one is able
to identify the endpoint of tachyon condensation and compute around
the correct vacuum. We will indeed claim  that the true vacuum 
above the Hagedorn temperature  has a deformed Poincar\'e symmetry
stemming from a discrete lattice structure in euclidean time, and that
the  divergence of $\L$ in the string computation
 is  a consequence of having
neglected this discreteness.

\subsection{An unexpected duality and time (de)construction}\label{unexp}

A  surprising result emerges looking in more detail into the
structure of the deformed algebra which corresponds to discrete
euclidean time. In this case the spacetime consists of a stack of
spacelike surfaces, separated by a spacing $a$ in euclidean time,
see fig.~\ref{fig0}, and on each spacelike surface are defined the
angular momentum operators $J_{ij}$, the position operators $x_i$ and
the momenta $p_i$. As we have seen, the $J_{ij}$ satisfy the
undeformed commutation relation of $SO(d)$,
\be\label{JJ}
[J_{ij}, J_{kl}]=-i(\d_{il}J_{jk}+\d_{jk}J_{il}
-\d_{ik}J_{jl}-\d_{jl}J_{ik})\, .
\ee
The $x_i, p_i$ have the standard commutation relations of a vector
with $J_{ij}$, and furthermore we have found
\be\label{xx}
[x_i,x_j] = -ia^2J_{ij}\, ,
\ee
\be\label{xp}
\[ x_i,p_j\]  =  i\,\delta_{ij}\,\, \sqrt{1+a^2({\bf p}^2+m^2)}
\, ,
\ee
while the momenta commute,
\be\label{pp}
[ p_i, p_j]=0\, .
\ee
Now observe that, if we introduce the notation
\be
J_{0i}\equiv \frac{x_i}{a}\, ,
\ee
eq.~(\ref{xx}) becomes
\be
[J_{0i},J_{0j}]=-iJ_{ij}
\ee
which is the commutation relation between  boosts in a (undeformed)
Lorentz group. 
\EPSFIGURE[t]{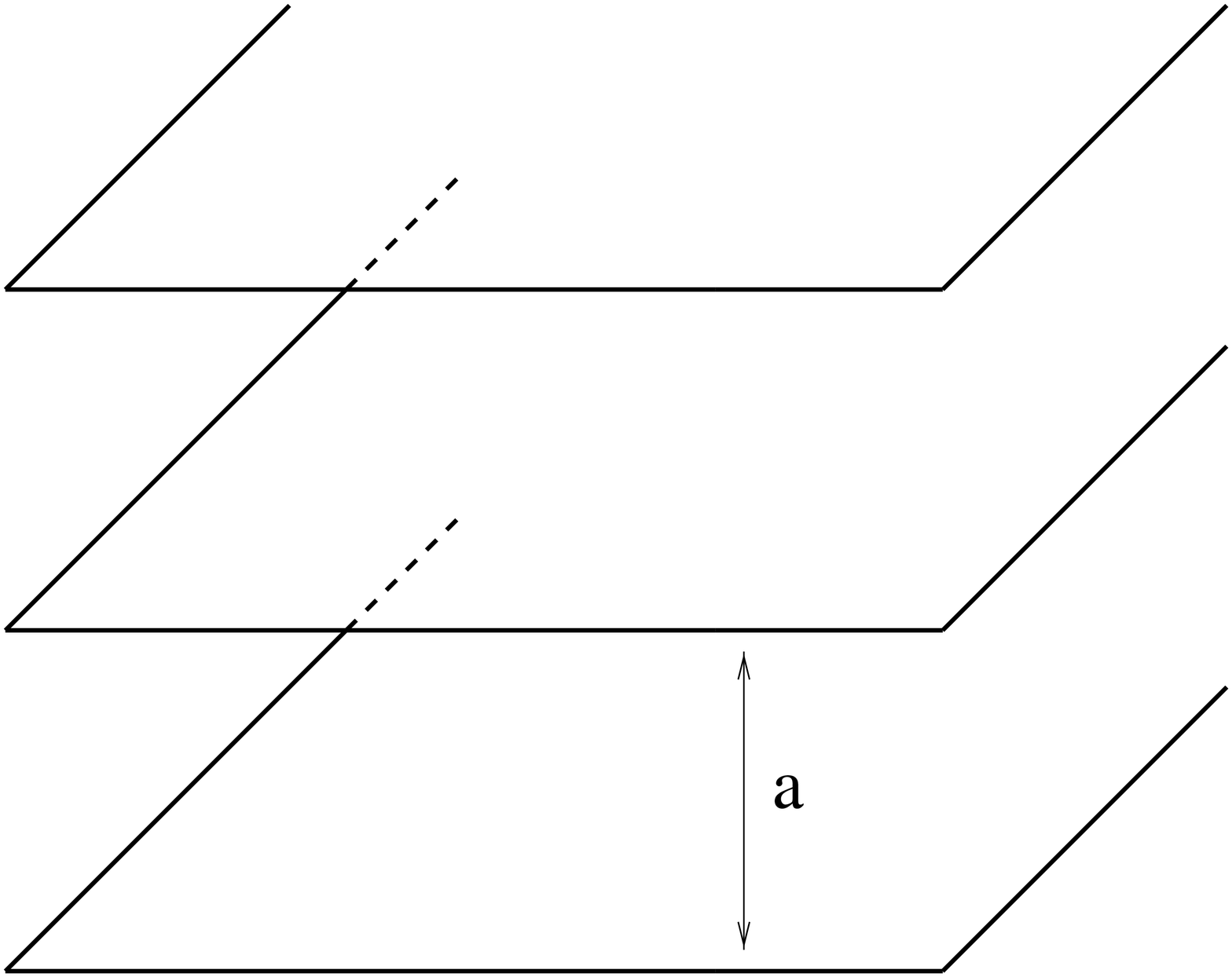, width=5cm}{The spacelike surfaces separated by
a distance $a$ in euclidean time.\label{fig0}}
Furthermore, since the $x_i$ are vectors, so are
the $J_{0i}$, and therefore also the 
boosts-angular momenta commutators $[J_{0i},J_{jk}]$ are
the same as in the Lorentz group. Therefore \eqs{JJ}{xx} reconstruct
a Lorentz group $SO(d,1)$. We can go even further observing that, in terms
of the variables $J_{0i}$, \eq{xp} reads
\be
[ J_{0i},p_j] =i\,\delta_{ij}\,\, 
\frac{1}{a}\sqrt{1+a^2({\bf p}^2+m^2)}=
 i\,\delta_{ij}\,\, \sqrt{{\bf p}^2+m^2+\frac{1}{a^2}}
\, .
\ee
However, this is nothing but the boost-momentum commutator of a
(undeformed) Poincar\'e group with Hamiltionian 
\be\label{H}
H= \[ {\bf p}^2+m^2+\frac{1}{a^2}\]^{1/2}\, .
\ee
This is just the Hamiltonian of a particle of mass $m^2+(1/a^2)$ and
therefore has automatically
also the correct commutation relations with
momenta and angular momenta, $[p_i,H]=[J_{ij},H]=0$, 
as well as with boosts, $[J_{0i},H]=ip_i$. Therefore,
together with \eq{pp}, we have reconstructed a full undeformed
Poincar\'e symmetry in $d+1$ spacetime dimensions! This is a truly
surprising result, since our starting point was a 
spacelike $d$-dimensional
surface with no time direction, with a non-commutative spatial
geometry on it, and we now discover that this can be interpreted as a $d+1$
dimensional commutative spacetime. We see that the
non-commutative spatial geometry has 
generated  a timelike dimension, somewhat similarly to the
(de)construction of a dimension discussed in ref.~\cite{AHCG}
(and generalized to a time dimension in ref.~\cite{BGK};
see also ref.~\cite{Ali} for a related approach based on non-commutative
geometry). 

It is clear that the physical interpretations of  
$x_i$ either as a position operator 
or as $a$ times a boost, are dual to each other. In
the limit of  $a$ small compared
to all other scales ($a|{\bf p}|\ll 1, am\ll 1$)
the non-commutativity of the
$x_i$ is small, and it is appropriate to interprete them
as coordinates on a manifold. At the same time the lattice spacing
between the spacelike sheets in fig.~\ref{fig0} goes to zero and a
continuous time is recovered. Instead when $a\ra 0$ \eq{H}
gives $H\simeq 1/a\ra\infty$, so that the extra time dimension
with respect to which $x_i/a$ is a boost becomes unaccessible.

The opposite situation takes place in the limit $a\ra\infty$. In this
case the separation between the spacelike sheets 
in fig.~\ref{fig0} goes to infinity and we are apparently left with a
single spatial surface, infinitely separated in euclidean
time by all others, so
the original
time has disappeared. However, in this limit the
non-commutativity of the $x_i$ becomes infinitely strong and it makes
no sense to interprete them as coordinates of a manifold. The correct
description is now in terms of boosts operators, and 
we recover a full (undeformed)
Poincar\'e symmetry, and therefore a new minkowskian time variable has
emerged.  Furthermore, in the limit $a\ra\infty$
the Hamiltonian in
\eq{H} becomes our original starting point before deformation,
$H=\sqrt{{\bf p}^2+m^2}$. Therefore at $a=\infty$ we recover, in an
unexpected way, the symmetries of the
undeformed theory with $a=0$.

\section{Application to string theory above the 
Hagedorn temperature}\label{sect4}

\subsection{The Atick-Witten free energy}\label{sect2}

We find useful to  recall in this section
some  known facts about the 
closed string free energy at finite temperature, and in particular 
its properties under T-duality.
In the closed bosonic string
the contribution  to the torus partition function of a single
string coordinate $X$ with a periodic identification $X\sim X+2\pi R$
is~(see e.g. ref.~\cite{Pol}, sect.~8.2)
\be\label{Z}
Z_X(\tau ,R)=|\eta (\tau )|^{-2}\sum_{n,w=-\infty}^{\infty}
\exp\[ -\pi\tau_2 
 \( \frac{\alpha 'n^2}{R^2}+\frac{w^2R^2}{\alpha '} \)
+2\pi i\tau_1nw \]\, ,
\ee
with $\tau$ the modular parameter of the torus and $\eta$ the Dedekind
eta function. With a Poisson resummation this can be written as
\be\label{Z2}
Z_X(\tau ,R)=
2\pi R\, Z_X(\tau )\sum_{m,w=-\infty}^{\infty}
\exp\( -\frac{\pi R^2 |m-w\tau |^2}{\a'\tau_2}\)\, ,
\ee
where $Z_X(\tau )$ is the partition function of the
non compact theory. Eq.~(\ref{Z}) shows explicitly the T-duality
symmetry  $R\ra\a '/R, n\leftrightarrow w$,
while \eq{Z2} shows explicitly the modular invariance: $\tau\ra \tau
+1$ is compensated by a change of variable $m\ra m+w$, and $\tau \ra
-1/\tau$ by $m\ra -w, w\ra m$.

Eqs.~(\ref{Z}, \ref{Z2})  hold also when the periodic field is
$X^0$, in which case we are studying  finite temperature   with
$T=1/(2\pi R)$ and, 
after including the contribution of all
the spatial  $X$'s and of ghosts
and integrating over the fundamental
domain of the torus, one gets $-F_1(T)/(VT)$,  where  $F_1(T)$ is
the one-loop free energy and
$V$ is the spatial volume. Taking explicitly the large
$T$ limit, Atick and Witten find the result 
\be\label{L1}
F_1(T)\ra \L_1 \, VT^2 \, ,\hspace{15mm} (T\ra\infty )
\ee
where $\L_1$ is the (divergent) one-loop cosmological constant of the bosonic
string. To get \eq{L1} it is necessary  to take 
the large $T$ limit inside the integral over the  moduli
space, and then one can replace the summation over windings with an integral
over continuous variables. 
Of course, strictly speaking these manipulations cannot be justified,
since they are performed on a divergent integral. However, comparison
with field theory shows~\cite{AW} that this interchange is dangerous only near
the corners of moduli space which correspond to UV regions, 
like $\tau \ra 0$ for the torus. Since these regions are absent in
string theory, there are good reasons to believe that \eq{L1} catches
the correct $T$ dependence in the large $T$ limit.\footnote{The
conclusion of ref.~\cite{AW} was indeed criticized  in
ref.~\cite{AK}, where it was remarked that the result is in
contradiction with the $T$-duality of the heterotic string,
and the blame was put on this exchange between the large $T$ limit and
the integration. We will
see however that at finite temperature
$T$-duality along $X^0$ {\em is} broken by 
world-sheet vortices above the Hagedorn temperature.}

Another way to get the same result is to use the invariance
under T-duality of  \eq{Z}.
Then, in units of the self-dual
temperature $T_{\rm self\, dual}=(2\pi\sqrt{\a'})^{-1}$, we have 
$F(T)/T =TF(1/T)$ so that, at large $T$,
$F(T)\simeq T^2F(0)$~\cite{AW,Pol}. However, while $Z_X(\tau ,R)$ at
fixed $\tau$
is certainly T-dual, 
as we see  from the explicit and well-defined expression (\ref{Z}), the
T-duality of $F(T)/T$ is  rather formal, since
$F(T)/T$ is obtained performing the integral over  the fundamental
domain of the torus of a T-dual integrand, but this integration
diverges at $\tau_2\ra\infty$
because of the tachyon. We will come back to this point later.

In the bosonic string many finite temperature issues are obscured by
the presence of the zero temperature tachyon.
It is therefore more instructive to consider type II strings.
The great difference in superstring theory comes  from the
GSO projection. The important point here is that 
when a  coordinate $X$ is compact and we use periodic boundary
conditions both on spacetime bosons and fermions (so that spacetime
supersymmetry is preserved)
the GSO projection is the
same as in the non-compact space.
However, when we impose periodic boundary conditions on spacetime
bosons and antiperiodic on spacetime
fermions (which is the case for instance when $X=X^0$
and we study finite temperature)
 the GSO projection is
different~\cite{AW,Rohm}, and there are additional minus signs which
depend on the windings.
The one-loop free energy with this GSO projection is~\cite{AW}
\be\label{F0}
\frac{F_1}{VT}=\frac{2\pi R}{16}\(\frac{1}{4\pi^2\a'}\)^5
\int_F \frac{d^2\tau}{\tau_2^6}|\eta (\tau )|^{-24}
\sum_{m,w=-\infty}^{\infty}Z_{f}(\tau ;m,w)
\exp\( -\frac{\pi R^2 |m-w\tau |^2}{\a'\tau_2}\)\, ,
\ee
where $Z_{f}(\tau ;m,w)$ comes from the contribution of the
world-sheet fermions and is a
 combination of theta functions 
(see eq.~(5.20) of ref.~\cite{AW}). A crucial point is that
$Z_{f}(\tau ;m,w)$ depends on $m, w$ because of the modified GSO
projection. Modular invariance is  respected, and in fact the GSO
projection has been fixed just requiring it. To examine T-duality,
we rewrite  eq.~(5.20) of ref.~\cite{AW} using 
the Poisson resummation
$\sum_m\exp\[ -\pi (m-b)^2/a\]=a^{1/2}\sum_n
\exp\( -\pi an^2+2\pi i bn\)$ and we get
\bees
&&\frac{F_1}{VT}=\frac{\pi\sqrt{\a'}}{16(4\pi^2\a' )^5}
\int_F\frac{d^2\tau}{\tau_2^{11/2}}\, |\eta (\tau )|^{-24}
\sum_{n,w=-\infty}^{\infty}\nonumber\\
&&\left\{ 
e^{ -\pi\tau_2  \( \frac{\alpha 'n^2}{R^2}+
                   \frac{R^2w^2}{\alpha '} \)
+2\pi i\tau_1wn }\,
\[ |\vartheta_2|^8+|\vartheta_3|^8+|\vartheta_4|^8 -
  e^{i\pi w} (\vartheta_3^4\bar{\vartheta}_4^4+
              \bar{\vartheta}_3^4{\vartheta}_4^4) \]+
\right.\label{F1}\\
&& 
+ \left.
e^{ -\pi\tau_2  \( \frac{\alpha '(n-1/2)^2}{R^2}+
                   \frac{R^2w^2}{\alpha '} \)
+2\pi i\tau_1w(n-1/2) }
\[ e^{i\pi w}
   (\vartheta_2^4\bar{\vartheta}_4^4+\bar{\vartheta}_2^4{\vartheta}_4^4)
 - (\vartheta_3^4\bar{\vartheta}_2^4+
              \bar{\vartheta}_3^4{\vartheta}_2^4) \]
\right\}
\, . \nonumber
\ees
where $\vartheta_i(\tau ) =\vartheta_i(0|\tau )$ are the Jacobi 
theta functions. 
We see that the dependence of $Z_f$ on
$m,w$ has generated a more complicated dependence on $n,w$,
and the above expression is not symmetric under
$R\ra\a'/R$, $n\leftrightarrow w$. 
As a result, {\em T-duality in the temporal direction 
is broken, and the origin of
this breaking is in the modified GSO projection}~\cite{AW}.
The fact that the symmetry between the momentum and winding modes is
broken by the GSO projection can  also be seen  directly on
the spectrum.
Consider for instance the winding and momentum modes of the
tachyon, $|0;n,w\rangle$. In the sector $w=0$ all momentum modes
 $|0;n,w=0\rangle$ are eliminated by the GSO projection, for all
$n$ (including of course the zero temperature tachyon $n=0$). 
Instead, in the sector $n=0$,  the winding modes
$|0;n=0,w\rangle$ with $w$ odd survive. 
The mass formula for the momentum
and winding modes of the tachyon in type II strings is
\be\label{2mass}
\a'm^2=-2+\frac{\a'}{R^2}n^2 +\frac{R^2}{\a'}w^2\, ,
\ee
so we see that at low temperature the lowest lying momentum modes 
of the tachyon with $w=0$ would have been themselves
tachyonic, so it is very welcome that they are all eliminated by the GSO
projection.\footnote{Of course, \eq{2mass} gives a mass in the remaining
nine-dimensional 
euclidean space, since we have separated the effect of the momentum
and winding in the
compact direction. To really interprete it as a mass of a state in
Minkowski spacetime we should rotate back the theory to Minkowski
along one of these nine directions rather than along $X^0$, 
and keep $X^0$ as a compact spatial coordinate. 
However, this 
is an issue of interpretation which is quite irrelevant. The real
point is that, when $m^2$ in \eq{2mass} is negative, there are
divergent contributions to the partition function, exactly as if there
were a tachyon in Minkowski space.}

The effect of the GSO projection can also be checked 
 expanding the integrand in 
\eq{F1} for $\tau_2\ra\infty$.
Recall that the spectrum of a string theory in D spacetime dimensions
can  be read off 
the asymptotics of the torus partition function at $\tau_2\ra\infty$,
which has the general form~\cite{Pol}
\be
\sim \int^{\infty} d\tau_2\,\tau_2^{-1-(D/2)}\sum_i\exp\(
 -\pi\tau_2\a' m_i^2\)\, ,
\ee
where $m_i$ are the masses of the states in the theory.
Expanding the integrand of \eq{F1} and retaining the terms
corresponding to the momentum and winding modes of the tachyon, one finds
that the asymptotic behavior at large $\tau_2$ is
\be
\sim
\int^{\infty} d\tau_2\, \tau_2^{-11/2}\, \sum_{n,w=-\infty}^{\infty}
\[ 2+ 480\, e^{-2\pi\tau_2}- (-1)^w (2-32\,  e^{-2\pi\tau_2}) \]
e^{ -\pi\tau_2 ( -2+\frac{\a'}{R^2}n^2 +\frac{R^2}{\a'}w^2 )}\, .
\ee
We see that, because of the factor $(-1)^w$, 
the winding modes of the tachyon with $w$ even are
eliminated from the spectrum
while those with $w$ odd survive\footnote{I thank
Maria Alice Gasparini for performing this check.}
(the prefactor corresponds to $D=9$ 
because we have separated the effect of the compact
direction to obtain an effective mass in the remaining 9 
dimensions). The winding modes at large $R$ are heavy, so  at low temperatures
the spectrum is free of tachyonic instabilities. As we decrease $R$
below a critical value $R_{\rm Hag}$, however, the two states
$|0;n=0,w=\pm 1\rangle$
become tachyonic. From \eq{2mass}, this happens at
\be
R_{\rm Hag}= \sqrt{2\alpha '}\, \hspace{5mm}\Rightarrow\hspace{5mm}
\thag =\frac{1}{2\pi\sqrt{2\a'}}\hspace{10mm}{\rm (type\,\,II)}
\ee
which is indeed the Hagedorn temperature of type II strings. 
The Hagedorn
transition is therefore signalled by the fact that, in an otherwise
tachyon free theory, a winding mode which is not removed by the GSO
projection becomes tachyonic above $\thag$ \cite{Sat,Kog,OBT,AW}. 

For the bosonic string again the states $w=\pm 1$ become tachyonic
exactly at its Hagedorn temperature, 
\be
R_{\rm Hag}= 2\sqrt{\alpha '}
\, \hspace{5mm}\Rightarrow\hspace{5mm}
\thag =\frac{1}{4\pi\sqrt{\a'}}\hspace{10mm}{\rm (bosonic)}\, .
\ee
However in the bosonic string the meaning of this finite temperature
tachyon is
obscured by the fact that the theory is already tachyonic at zero
temperature, and that there are also all momentum modes of the tachyon,
which instead switch from tachyonic to non-tachyonic as we decrease $R$.

Using \eq{F0} and
performing explicitly the large $T$ limit of the one loop free energy,
Atick and Witten  find also for type~II strings
the result $F_1(T)=\L_1 VT^2$. Thus,
the  $T^2$ dependence of the free energy  still appears,
but it is not anymore a  consequence of T-duality, which now is
explicitly broken.
Indeed, taking the  small $R$ limit of type~IIB
theory we do not find type~IIA at large $R$, as would be the case for
the compactification in the absence of the $(-1)^{\bf F}$ twist, 
but rather type~0 theory  \cite{AW}.  The spacetime supersymmetry
of type II strings is broken by the
$(-1)^{\bf F}$ twist in the boundary conditions; fermions have
 half-integral momenta while boson have integral momenta. In the limit
$R\ra 0$ all fermions are therefore removed from the spectrum, and we
are left with the bosonic spectrum of the type~0A theory at $R\ra 0$.
The tachyon that develops
at the Hagedorn transition becomes, in the  limit $R\ra 0$,
the tachyon of the type~0
theory.

In the large $T$ limit the quantity
$\L_1$, computed from string theory around this tachyonic vacuum,
 is  given by the  type~0 partition function~\cite{AW}, 
\be\label{L1type0}
\L_1 =\frac{1}{16}\( \frac{1}{4\pi^2\a'}\)^4
\int_F\frac{d^2\tau}{\tau_2^6}|\eta (\tau )|^{-24}
\[ |\vartheta_2|^8 + |\vartheta_3|^8 +
|\vartheta_4|^8 \]\, .
\ee
The integration over the fundamental region of the torus
moduli space is
divergent at $\tau_2\ra \infty$ because of the type~0 tachyon.

Finally, the behavior that we have discussed so far is the free
energy at one-loop. The  contributions at $k$-loops have the
form~\cite{AW}
\be
F_k(T)\sim VT^2\, ( g^2T^24\pi^2\a' )^{k-1}
\ee
The factor in parenthesis is just $g^2\a'/R^2$. Therefore in the
region
\be\label{window}
g\sqrt{\a'}\ll R \ll \sqrt{2\alpha '}
\ee
the leading contribution to the free energy is the one-loop term and
$F(T)\sim VT^2$. At $R=g\sqrt{\a'}$ all higher loop contributions
become comparable and a change of regime takes place. Of course, we
know nowadays that what happens at this scale is 
that D-branes become important and, in type IIA theory, we see
the opening up of the
11th dimension. With hindsight, it is interesting to observe that the
importance of the mass scale $1/(g\sqrt{\a'})$ in string theory could
have been inferred already from the higher loop behavior of the free energy.

\subsection{World-sheet vortices and T-duality}\label{sect.ws}

The partition function of the closed bosonic string
at finite temperature  is computed working 
with the euclidean field $X^0(\sigma_1 ,\sigma_2)$
periodically identified, $X^0\sim X^0+2\pi R$.  
The dynamics of $X^0$
is governed by the action
\be\label{Scont}
S=\frac{1}{4\pi\a'}\int d^2\xi\,\, (\pa_{\a} X^0 )^2 \equiv
\b\int d^2\xi\,\, \frac{1}{2}(\pa_{\a} \theta )^2
\ee
where $X^0\equiv R\theta$ so that $\th\sim \th +2\pi$, 
and
\be
\b \equiv \frac{R^2}{2\pi\a'} \, .
\ee
There is however a crucial subtle
point. Because of the identification $\th\sim\th +2\pi$, this is
a non trivial theory and the continuum
definition (\ref{Scont}) is still somewhat formal. In fact,  
its possible discretizations  fall into two different
universality classes, depending on whether they admit or not vortices.
To elucidate this point, let us recall that
in the continuum limit
a vortex is a classical configuration  singular at one point, 
such that as we encircle the singular point once, the field $\th$ does
not come back to itself but rather to $\th +2\pi v$, with $v$ integer.
The form of this configuration is
$\th (\s_1 ,\s_2)=v\arctan (\s_2/\s_1)$ which, substituted into
\eq{Scont}, gives the vortex action
\be\label{Svortex}
S_{\rm vortex}=\frac{\beta v^2}{2}\int \frac{d^2\xi}{|\xi |^2}=
\pi\b v^2\log (L/\ep )\, , 
\ee
where we used a lattice discretization to regularize the integral in
the UV and a finite volume of the world-sheet to regularize it in the
infrared.\footnote{ 
Actually, on a lattice we cannot follow  $\th$ continuously as we
encircle a point, and
the definition of vortices must be modified.
We will come back later to the correct lattice definiton of vortices;
however for $\ep\ra 0$ these lattice vortices will reduce to the
continuum definition and for a first estimate we can use  \eq{Svortex}.}
$S_{\rm vortex}$ 
is therefore divergent as $\ep\ra 0$
and vortices might seem to be irrelevant
in the continuum limit.
However, they have a
collective coordinate which is the position of the center, so their
multiplicity is equal to the number of lattice sites and the
contribution to the partition function from the vortices with $v=\pm 1$ is
\be\label{Zcont}
Z_{v=\pm 1}\sim \(\frac{L}{\ep}\)^2 e^{-\pi\b\log (L/\ep )}=
e^{ (2-\pi\b )\log (L/\ep )}\, .
\ee
We see that for $\beta >2/\pi$ vortices are indeed irrelevant, but for
$\beta <2/\pi$  vortices with $v=\pm 1$
dominate; this is the famous Kosterlitz-Thouless (KT) phase
transition~\cite{Ber,KT,Kos};
$\beta_c =2/\pi$ corresponds to 
$R=2\sqrt{\alpha '}$ and therefore to $T=1/(4\pi\sqrt{\a'})$, which
coincides with the Hagedorn temperature of the bosonic
string~\cite{Sat,Kog}. The Hagedorn phase transition in spacetime
is therefore signalled by a KT transition on the world-sheet.

The same analysis can be repeated for the supersymmetric KT
transition, and one finds now 
a critical value $\beta_c =1/\pi $~\cite{Gol,Sat},
that corresponds to $T=1/(2\pi\sqrt{2\a'})$, so that 
also for type~II strings (and for the heterotic string~\cite{OBT})
we recover the correct value of the Hagedorn temperature.

It is however also possible
to find  different discrete formulations in
which vortices are explicitly suppressed, 
and in this case one finds that the KT transition is absent~\cite{GK1}.
The question  is which of the two types of
discretizations should be taken as fundamental,
that is, which of the two is the correct definition for the
formal expression~(\ref{Scont}), in the application to string theory.
A clue to this
issue is the fact that vortices and the KT transition
break T-duality. This is clear from 
the fact that in the large $\b$ phase the theory has 
 an infinite correlation length while in the low $\b$ phase a
standard lattice strong coupling expansion shows that a finite correlation
length is generated.
If however one explictly suppresses the
vortices, one finds that the KT transition is eliminated and T-duality
is restored~\cite{GK1}.

The answer that we propose is therefore the following:
when the compactification of a superstring is done along a  circle
with  periodic boundary conditions on bosons and fermions,
we know that 
T-duality is respected, and therefore the definition that suppresses the
vortices is the correct one.
But when we compactify $X^0$ to study  finite temperature, 
or more in general when we compactify a coordinate on a circle
with a  $(-1)^{\bf F}$ twist,
the situation is different.
In sect.~2 we have seen that in type~II theory thermal duality 
$T/T_{\rm self\, dual}\ra T_{\rm self\, dual}/T$ {\em is}
broken.
Therefore when we compactify 
 with a $(-1)^{\bf F}$ twist we
{\em must} chose a discretization which retains the vortices. 
The fact that we have two different prescriptions for the definition
of \eq{Scont} on $S^1$ and on $S^1/(-1)^{\bf F}$
goes back to the fact that in these
two cases we have also two different prescriptions for the GSO projection.

In the bosonic string instead  the question of whether  thermal duality
$F(T)/T=TF(1/T)$ really holds is not well posed, since $F(T)$ diverges
at all temperatures because of the tachyon.
In the bosonic string
one has the tendency to ask questions forgetting about the tachyon,
since this will be cured by the superstring. However, the 
mechanism that in the superstring
eliminates the zero temperature tachyon and, at finite
temperature, all its tachyonic 
momentum modes, is the GSO projection, which is
the same mechanism that breaks $T$-duality in the temporal direction. 
Therefore
the two issues cannot be separated.
However, as long as we
wish to use the bosonic string as a simplified model 
to learn something about
superstrings, we must also use the same prescription. In the following
analysis we will  use for simplicity the bosonic string,
retaining the vortices when compactifying $X^0$.

Similarly, using this prescription on the field $X^0$ of the heterotic
string, one finds again a KT phase transition on the world-sheet at a
value of $R$ corresponding to the Hagedorn temperature, and above
$\thag$ T-duality is broken. The need for a mechanism that breaks
thermal duality in the heterotic string was discussed in ref.~\cite{AW}.

\subsection{Decompactification and
dynamical localization of $\tilde{X}^0$}\label{sect4.2}

Let us therefore recall in more detail what is the effect of vortices,
following the review~\cite{Kogut}. 
Consider  a lattice discretization of \eq{Scont}, 
so that the partition function reads
\be\label{Se}
Z=\int [\prod_{r}d\th (r)]\, \exp\left\{
-\beta\,\sum_{r,\a } \, \frac{1}{2} (\D_{\a} \theta )^2 \right\}
\, ,
\ee
where
 $\D_{\a}\th (\xi) =(\th (\xi +\ep \hat{\a})
-\th (\xi -\ep \hat{\a}))/2$, $r$ is an index labelling the lattice sites
and $\ep$ is the world-sheet lattice spacing, to be eventually sent to
zero.\footnote{As $\eps\ra 0$,
$\D_{\a}\ra\ep\pa_{\a}$ to conform with
the notations of ref.~\cite{Kogut}.}

With standard manipulations (see e.g. ref.~\cite{Kogut},
eqs.~(7.19)-(7.28), or ref.~\cite{JKKN}) it can be rewritten as
\be\label{ZZ}
Z=   \int [\prod_{r} d\f (r) ] \sum_{ \{m(r)\} =-\infty}^{\infty} 
\exp\left\{ -\frac{1}{\b}\sum_{r,\a} \frac{1}{2}(\D_a\f )^2
+2\pi i \sum_r m(r)\f (r)\right\}\, .
\ee
Here $\f$ is a real scalar field that, contrary to $\th$, 
{\em is  not subject to any  periodic identification}, 
i.e. $\f$ lives on $R$ rather than on $S^1$,
and
$m(r)$ is 
an integer valued field.\footnote{This is a particular case of
a more general construction~\cite{AbK}:  if $N$ is a non-simply
connected manifold, $M$ its universal covering space and $N=M/G$
with $G$ a group freely acting on $M$,
 then a field on $N$ can be similarly replaced
by a field on $M$ plus a field in $G$; here
$S^1=R/Z$, so $\f\in R$ and $m\in Z$.
When $G$ is a continuous group the field in $G$ is a gauge field.}
At small $\b$, when $\th$ is strongly coupled, $\f$ is instead weakly
coupled and viceversa. Thus $\phi$ is the most convenient variable
 in the region $\b <2/\pi$, i.e $T>\thag$, where it
 describes an ordinary massless scalar field, weakly coupled and
unconstrained.
The integer valued field $m(r)$ describes instead the
vortices. This can be shown integrating out the $\f$ field;
then one finds the vortices partition function
\be\label{Zv}
Z_{\rm vortices}\sim \sum_{ \{ m(r)\} =-\infty }^{\infty}\exp
\left\{ -2\pi^2\b \sum_{r,r'} m(r)G(r-r')m(r') \right\}\, .
\ee
$G(r-r')$ is the lattice propagator, $\D^2G(r)=\d_{r,0}$, and 
at $r=r'$  it diverges  logarithmically, $G(0)\simeq (1/2\pi)\log
(L/\ep)$. Therefore the term with $r=r'$ in the sum in
\eq{Zv}  gives a contribution to the
action
\be
S (\{ m(r) \})=
\pi\b \(\sum_r m(r)\)^2 \log (L/\ep )\, .
\ee
Comparison with \eq{Svortex} shows that the configuration with 
$m(r)=v\delta_{r,r_0}$, with $v$ integer, 
can be identified with a vortex centered at $r_0$
and with winding $v$, and
this provides the  lattice definition of the vortex.
The contribution of $G(r-r')$ at $r\neq r'$ provides instead an
interaction term between the vortices, which grows logarithmically
with $r-r'$.

For our purposes it is more useful to integrate out the vortices and
remain with an effective field theory for $\f$. We follow again
refs.~\cite{Kogut,JKKN}. Because
of the long-range interaction between vortices, this is a non-trivial
many body problem that can be addresses using the renormalization group
(RG). A RG transformation of the partition function
(\ref{ZZ}) generates also a term
$\sim \sum_r m^2(r)$ in the action, so it is convenient to start
directly with
\be\label{ZZZ}
Z=   \int [\prod_{r} d\f (r) ] \sum_{ \{m(r)\} =-\infty}^{\infty} 
\exp\left\{ -\frac{1}{\b}\sum_{r,\a} \frac{1}{2}(\D_a\f )^2
+(\log y)\sum_r m^2(r)
+2\pi i \sum_r m(r)\f (r)\right\}\, .
\ee
The initial value for the RG transformation is $y=1$. 
Defining $x=\pi\b -2$,
the RG flow in the $(x,y)$ plane turns out
to have a line of fixed points at $y=0, x\geq 0$.
At $y$ close to zero \eq{ZZZ} simplifies because we can retain only
the states with $m=0,\pm 1$. Using 
\be
 \sum_{m(r)=0,\pm 1}
e^{\{ (\log y) m^2(r)+2\pi i m(r)\f (r) \}}=
1+2y\cos\[ 2\pi\f (r)\]
\simeq \exp\{ 2y\cos\[ 2\pi\f (r)\]\}\, 
\ee
and rescaling $\f\ra \f\sqrt{\b}$, \eq{ZZZ} becomes
\be\label{SG}
Z\simeq   \int [\prod_{r} d\f (r) ] 
\exp\left\{ -\frac{1}{2}\sum_{r,\a} (\D_a\f )^2
+2 y\sum_r\cos\[ 2\pi\sqrt{\b}\f (r)\]
\right\}\, ,
\ee
and we recognize a sine-Gordon theory. 

\EPSFIGURE[t]{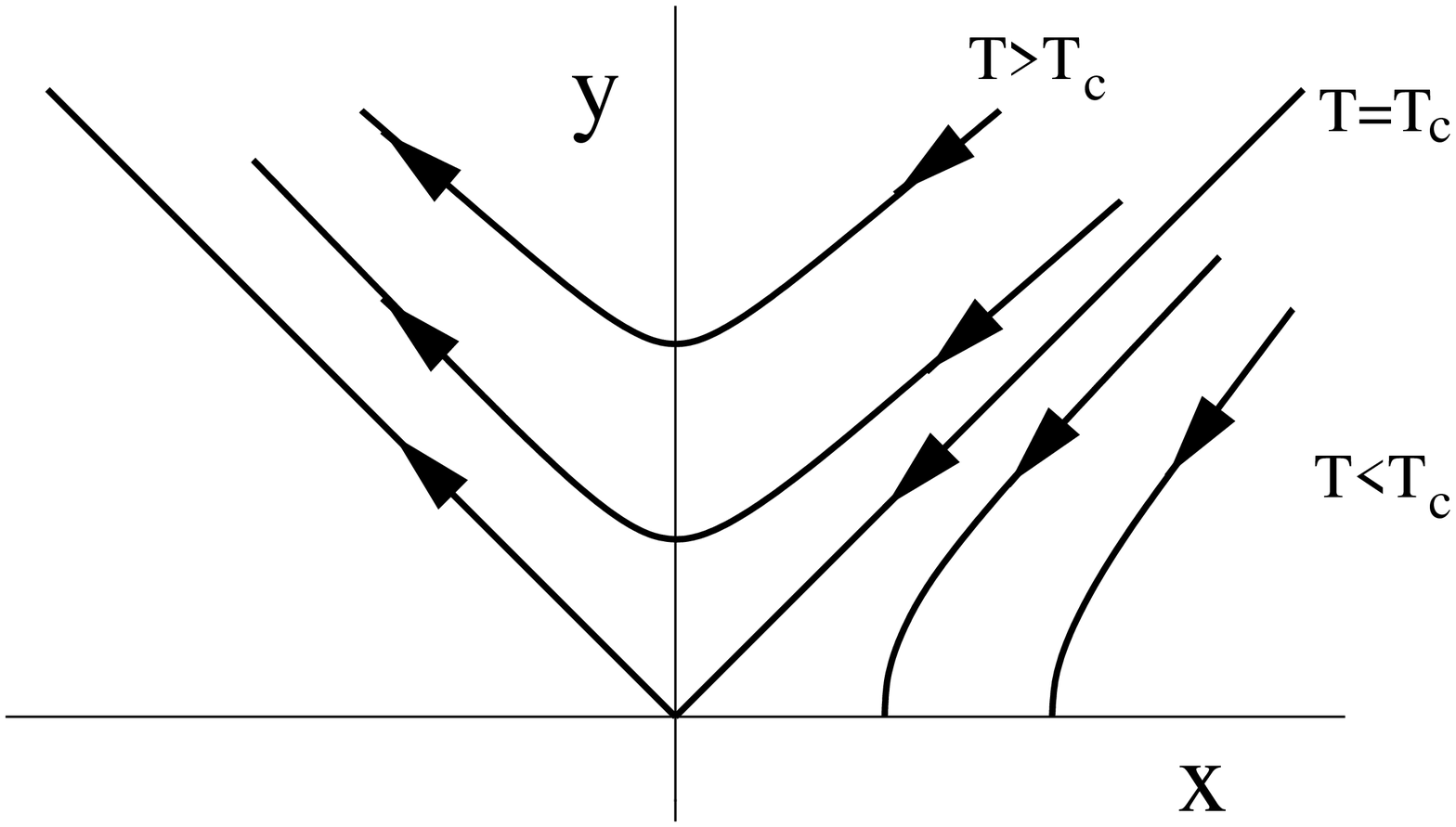, width=5cm}{The RG flow in the (x,y) plane.
\label{fig2}}
\vspace*{5mm}
%\FIGURE[t]{\epsfig{file=fig1.eps}{The RG flow in the (x,y) plane.}}
%\begin{figure}
%\centering\leavevmode\epsfxsize=2.5in\epsfbox{fig1.eps}
%\caption{The RG flow in the (x,y) plane.
%\label{fig1}
%}
%\end{figure}

The RG flow  in the $(x,y)$
plane is given by the well-known 
diagram reproduced in fig.~\ref{fig2}~\cite{KT,Kos,Kogut} (and the general
KT picture has been  confirmed by many Monte Carlo simulations
and lattice strong
coupling expansions, see e.g. refs.~\cite{Gupta}--\cite{Campo} and
references therein, while the existence of the phase transition was rigorously
proved in ref.~\cite{FS}).  
From fig.~\ref{fig2} 
we see that there is a critical value  $\b_c$ and therefore
  a critical temperature $T_c$ (in a first
approximation $T_c=\thag$, but we will come back to this point below),
such that for $T< T_c$ the RG trajectories flow toward $y=0,
x>0$ and then stop, i.e. we have a line of fixed points. 
At $y=0$ vortices are completely suppressed, as we read from
\eq{ZZZ}, and because of this the issue of whether to include or not
vortices in the regularized theory is irrelevant 
in the low temperature phase. The critical properties are the same as
those of a free scalar field, and the correlation length is infinite.

Above $T_c$,
however, $y$ flows toward large values and vortices are important.
In this regime, we have seen that the weakly coupled
variable is $\f$ rather than $\th$. We therefore 
 describe the string at $T>T_c$ by
${\tilde{X}^0}$ and the $X^i$, where
\be
\tilde{X}^0\equiv \sqrt{2\pi\a'}\,\, \f\, .
\ee
The normalization of $\tilde{X}^0$ has been chosen so that its kinetic
term has the standard string normalization
(we will discuss in sect.~\ref{sect.tt} the relation of 
$\tilde{X}^0$ to $X_L^0-X_R^0$). Observe that $\f$ is not subject 
to any periodic identification, and therefore
we have no periodic identification on ${\tilde{X}^0}$
either: the domain of definition of ${\tilde{X}^0}$ is the whole real
line, rather than a circle.
Since the effect of
vortices has already been taken into account, we can use without
ambiguity a continuum notation, and the action for
$\tilde{X}^0$ reads
\be\label{SGX}
S=
\int d^2\s\,\,\[
\frac{1}{4\pi\a'} (\pa_a \tilde{X}^0 )^2
-\mu \, \cos\( \frac{R}{\a'}\,\tilde{X}^0 \)\]
\, ,
\ee
with $\mu =2y/\eps^2$.
Therefore  below $\b_c$ we have 
an effective potential for ${\tilde{X}^0}$
\be\label{pot}
V({\tilde{X}^0})=-\mu \cos\( \frac{R}{\a'}\,\tilde{X}^0 \)\, ,
\ee
with $\mu\ra \infty$.
This potential
breaks the continuous translation symmetry of ${\tilde{X}^0}$ to discrete
translations, and localizes ${\tilde{X}^0}$ on the minima of the cosine, i.e.
on an infinite lattice with spacing
$a= 2\pi\a'/R$.

The  same results on the decompactification and dynamical localization
of the compact coordinate below $R_c$
that we have read off this well-known RG analysis of
the KT transition have also been obtained some time ago by Gross and
Klebanov~\cite{GK1,GK2} with matrix model techniques, and their result
\cite{GK2} is indeed that below $R_c$ the model defined by a single
compact string coordinate coincides with an infinite set of decoupled
$c=0$ one matrix models (see also refs.~\cite{Kut,HK}).\footnote{It is
also interesting to compare with the case of a boundary sine-Gordon
theory, corresponding to the insertion of the vertex operator of {\em
open} string tachyons. In this case the model is exactly solvable by
Bethe ansatz techniques~\cite{FSW} and one finds that the
boundary value of $X$ is pinned to the minima of the potential,
so that the IR fixed point is a stack of D-branes~\cite{HKM}.}

At the critical temperature the lattice spacing 
jumps from zero to the finite value
\be\label{aaa}
a_c=\frac{2\pi \a'}{R_c}\simeq \pi\sqrt{\a'}\, .
\ee
Therefore $a$ is the order parameter of a first order phase
transition in spacetime. The fact that the transition 
is first order is in agreement with the result
of ref.~\cite{AW}.

The value of the critical temperature $T_c$ is not exactly
equal to $\thag$, i.e. $\b_c$ is not exactly equal to $2/\pi$, as was
suggested by the  result~(\ref{Zcont}). 
The RG analysis \cite{KT,Kos,Kogut} shows that the critical
value of $\b$ is determined by
\be
\pi \b_c-2 = (2c)\, \exp\left\{-\frac{\pi^2}{2}\b_c\right\}\, ,
\ee
with $c\simeq  1.3\pi$ a  positive constant.
Physically, the correction term comes from the interaction between
vortices, while \eq{Zcont} was obtained in the diluite vortex approximation,
and the positivity of $c$ reflects  the formation of
an effective dielectric constant greater than one~\cite{KT}.
This gives a critical temperature
$T_c$ slightly {\em smaller} than $\thag =1/(4\pi\sqrt{\a'})$, 
\be\label{Tc}
\(\frac{\thag}{T_c}\)^2 = 1+ c\, 
\exp\left\{-\pi \(\frac{\thag}{T_c}\)^2  \right\}\, ,
\ee
which, solved numerically,  gives $T_c\simeq 0.94\thag$.
Eq.~(\ref{Tc}) is  compatible with a first order phase
transition in spacetime, and in particular with the fact that 
a first order transition
proceeds via tunneling before $\thag$ is reached;
denoting by $t$ and $t^*$ the winding modes of the tachyon with winding
$w=1$ and $w=-1$, respectively, the spacetime
dynamics is governed by an effective
potential of the form~\cite{AW}
\be
V(t^*t)=m^2(T)\, t^*t +u(T) (t^*t)^2+\ldots
\ee
where 
\be
m^2(T)=\frac{4}{\a'} \( \frac{\thag^2}{T^2}-1 \)
\ee
is the mass squared of $t,t^*$, and becomes zero at $\thag$; 
if $u(\thag )$ were positive, the transition could be second order
(depending also on the sign of the higher order terms) 
in which case it would take place when  $m(T)$ vanishes, i.e.
at $T =\thag$. However $u(\thag )$
turns out to be negative because of the tachyon-dilaton
coupling~\cite{AW}, and the transition is then first order and proceeds
via tunnelling when $m^2(T)$ is still positive, i.e. at $T_c<\thag$.

The fact that the transition is first order means that
the Hagedorn temperature is not limiting neither for
closed nor for open strings, since the transition takes place via
tunneling before $\thag$ is reached, independently of whether an
infinite energy would be needed to reach $\thag$
(which appears to be the case for open strings~\cite{DDGR}).

\subsection{Closed tachyon condensation}\label{sect.tt}

In the last few years there has been formidable progress in
understanding the condensation of  open string
tachyons~\cite{Sen}. For open strings, the vertex operators
of the tachyon live on the
boundary of the worlds sheet and, at small string coupling, their 
effect is just to
modify the boundary conditions of the world-sheet theory. Because of
this, the condensation of open string tachyons does not have a
dramatic influence on the structure of space-time itself. The typical
process that they describe is the annihilation of brane-antibrane
systems, whose endpoint turns out to be simply the flat vacuum.
It is instead believed that the condensation of closed string tachyon
is a much harder problem, because it involves a bulk perturbation of
the world-sheet. However, the results of the previous section 
give an answer to what happens when a closed string
tachyon with non-zero winding
condense. First of all, it is 
instructive to recover the same results with a
$\s$-model analysis of tachyon condensation. 

The vertex operator for the winding modes 
$w=\pm 1$ of the bosonic string tachyon is 
\bees
V (0,n=0 ,w=\pm 1)
&=&g_c\int d^2z\, :\exp\{ -i(k_L^0X_L^0+k_R^0X_R^0) +ik^iX^i\}:\nonumber\\
&=&g_c\int d^2z\, :\exp\{ \mp i\frac{R}{\a'}(X_L^0-X_R^0)
+ik^iX^i\}:\, ,
\ees
with $k_L=n/R+wR/\a' =\pm R/\a'$, $k_R=n/R-wR/\a' =\mp R/\a'$, and
the mass shell condition is ${\bf k}^2=(4/\a' ) (1-\thag^2/T^2)$, so that
$k_i=0$ just  at $T=\thag$.

The $\s$-model relevant for the  condensation of tachyons with $w=\pm 1$
is then 
\be\label{Ssigma}
S=\int d^2z\, \frac{1}{4\pi\a'}\( \pa_{\a}X^0 \)^2+ 
t(X)e^{  i\frac{R}{\a'}(X_L^0-X_R^0)} +
 t^*(X)e^{ - i\frac{R}{\a'}(X_L^0-X_R^0)} 
\, ,
\ee
where $t(X)$ is the complex tachyon field that describes the two
real tachyons with $w=\pm 1$. Observe that
the kinetic term of $X^0=X_L^0+X_R^0$ is equal to minus the kinetic
term of $X_L^0-X_R^0$, since $( \pa_{\a}X^0 )^2=2\pa_zX_L^0
\pa_{\bar{z}}X_R^0$; therefore, the action (\ref{Ssigma}) 
can be written  as a functional
of the combination $X_L^0-X_R^0$ only, and this is a peculiarity of a
tachyon which is a winding mode of the zero temperature tachyon, 
because it is just in this case that its vertex operator
depends only on $X_L^0-X_R^0$.
Writing $t(X)=|t|e^{i\vartheta}$, 
\eq{Ssigma} becomes
\be\label{Ssigma2}
S=-\int d^2z\, \frac{1}{4\pi\a'}\[ \pa_{\a}(X_L^0-X_R^0)\]^2
- 2|t(X)|\cos\[  \frac{R}{\a'}(X_L^0-X_R^0) +\vartheta \]
\, .
\ee
Comparison with \eq{SGX} shows that the two actions are the 
same\footnote{Neglecting an overall minus sign of the action. This has
no influence on the Noether charges of the theory and therefore on
spacetime quantities, since
in any case the sign of the Noether charges is unrelated to the sign
of the action, and is fixed by physical requirements.} 
if we take the tachyon field constant,
we identify $2|t|$ with $\mu$, and we set
\be\label{ide}
\tilde{X}^0=(X_L^0-X_R^0)+\frac{\a'}{R}\vartheta\, .
\ee
Actually, the identification (\ref{ide}) seems at first impossible because
we have seen that  $\f$, and hence 
$\tilde{X}^0$, is not subject to any periodic
identification, while we are used to the fact that,
 when a string coordinate $X=X_L+X_R$ is periodically
identified as $X\sim X+2\pi R$,  then
$X_L-X_R$ is  its T-dual variable and is
also periodically identified, with a periodicity $2\pi \a'/R$.
Since by definition also $\vartheta\sim \vartheta +2\pi$, this seems
to suggest that the right-hand side of \eq{ide}  lives on a circle of
radius $2\pi \a'/R$.

However, when $X=X^0$ (or in general when we compactify 
on $S^1/(-1)^{\bf F}$)
this conclusion is incorrect. The reason why
one usually says that $X_L-X_R$ is  periodic is the following. 
If $X=X_L+X_R$ is taken by definition to live on a circle of radius
$R$, then
the expansion of $X_L,X_R$ on the complex plane is~\cite{Pol}
\bees
X_L&=&x_L-i\frac{\a'}{2}p_L\log z +{\rm oscillators}   \nonumber\\
X_R&=&x_R-i\frac{\a'}{2}p_R\log \bar{z} +{\rm oscillators} 
\ees
with $p_L=(n/R)+(wR/\a'), p_R=(n/R)-(wR/\a')$.
Under a $2\pi$ rotation in the complex plane $X_L\ra X_L+\a'\pi p_L$,
 $X_R\ra X_R-\a'\pi p_R $ and therefore 
 $X_L-X_R\ra X_L-X_R+ n 2\pi R'$, with $R'=\a'/R$.
Now, {\em if} $X_L- X_R$ is a single valued
functions, we conclude that $X_L- X_R$ and $X_L- X_R +2\pi R'$
must be  identified, and  therefore also $X_L- X_R$ lives on a circle.

However, we have seen that when we compactify the time direction we
must allow for vortex configurations, and these are not single
valued. Below $\thag$ they are anyway dynamically irrelevant, so all
configurations of $X_L, X_R$ that contribute to the path integral are
single valued, and $X_L^0-X_R^0$ indeed lives on the dual circle.
Instead, above $\thag$ the path
integral is dominated just by the non single-valued configurations, and
therefore the above argument does not go through and 
$X_L^0- X_R^0$ has no constraint and
lives on the real line. The identification (\ref{ide})
is therefore possible, and
the sigma-model approach  gives the same answer as the
Kosterlitz-Thouless analysis  
 of the previous section.

\vspace{5mm}

We have therefore understood that, as we reach the Hagedorn
temperature from the low $T$ side, tachyon condensation leads us to a
state that can be described as a stack of spacelike surfaces,
separated by a spacing $a_c\simeq \pi\sqrt{\a'}$ in the euclidean time
direction, as in fig.~\ref{fig0}. 
Similarly, if we ask what is the endpoint of tachyon
condensation of a theory compactified at a radius $R<R_c$, we find a
stack of spacelike surfaces with $a=2\pi\a'/R$. The situation
is depicted in fig.~\ref{fig3}.

\EPSFIGURE[ht]{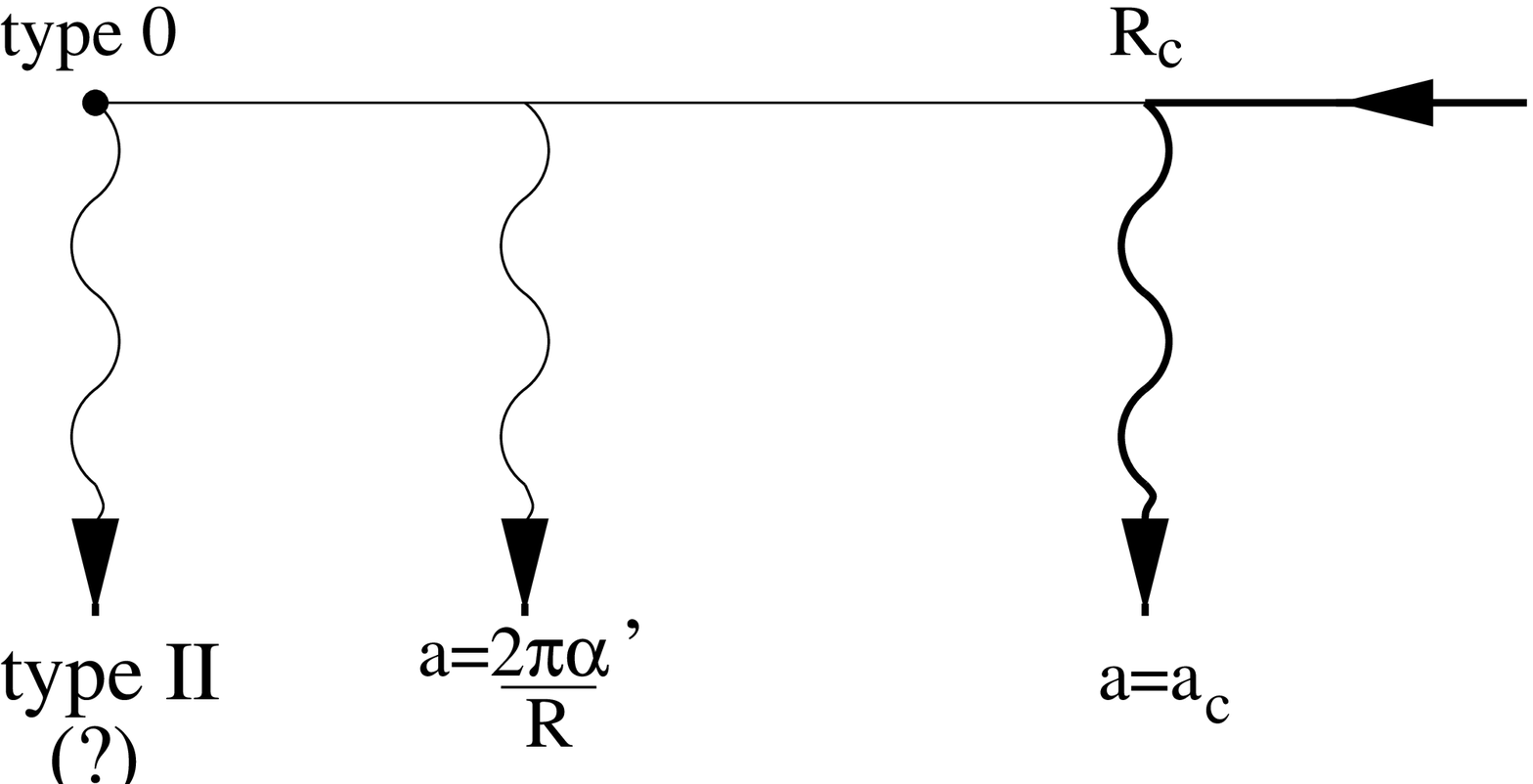, width=7cm}{Decreasing $R$ coming from the
region $R>R_c$ we end up with a lattice spacing $a_c$. The decay of
the theories with $R<R_c$ is also shown.
\label{fig3}}

The light degrees of freedom in the
new phase are naturally
identified with the fluctuation modes of these
spacelike surfaces. The peculiarity of these modes is that they evolve
in a space with a discretized euclidean time and, as we
have seen in sect.~\ref{sect3}, in this setting it is quite
natural to quantize them imposing the deformed commutation
relations (\ref{xxe}, \ref{xpe}). Therefore, if 
after reaching $\thag$ we inject further energy
into the system,  the free energy  of these modes, at large $T$, 
will have the Atick-Witten form $F(T)=V\L T^2$ with $\L$ finite and 
given by \eq{L1a}. 

\vspace{5mm}

We conclude this section with a few comments on the literature. First,
the fact that the  condensation of closed string
tachyons can produce
a discrete spacetime was nicely shown, in a different setting, 
in ref.~\cite{EFR}. These
authors considered string theory in D=2 (at zero temperature). 
In this case the would-be zero temperature 
tachyon is actually massless, but they put it slightly off-shell, 
$k^2=2-\eps$, so
that the tachyon operator becomes slightly relevant and it suffices
to compute the beta function at order $\eps$ to find the endpoint of
the condensation of this tachyon. Guided by the
analogy with Landau theory of solidification, they find that the final
spacetime is a two-dimensional lattice. This fits very nicely with our
results. 

Recently there have been a number of investigations on the 
condensation of type~0
tachyons~\cite{CG,GS,Adams,David}, and
signals of the formation of a lattice  structure in spacetime have 
also been found in some of these works.
In particular, Adams, Polchinski
and Silverstein~\cite{Adams} consider type~II theory in ten
dimensions, and replace a plane, say (89), with an orbifold
$C/Z_n$, i.e. identify $z\sim z\exp\{ 2\pi i/n\}$, where $z$ is the
complex coordinate of the (89) plane. If $z=re^{i\theta}$, at fixed
$r\neq 0$ the variable $r\th$ parametrizes a circle of length
$2\pi r/n$, so for given fixed $r$,  
in the  large $n$ limit we have a small circle. Furthermore 
in the large $n$ limit the orbifold
projection is such that going around this circle the fermions pick a
minus sign, i.e. we have a $(-1)^{\bf F}$ twist.
Taking the T-dual of this circle one has type~0 theory in the bulk,
and the authors find that 
{\em translation symmetry in the T-dual variable is broken}, and 
that there is a set of equally spaced branes
(defined broadly as defects which break translation invariance)
in the T-dual variable. This
is in agreement with the result that we have found. However, 
in our case we have also found that the would-be 
T-dual variable becomes non-compact, and lives on the
whole real line rather than on the dual circle. The difference might be
due to the fact that the tachyon studied in ref.~\cite{Adams} 
resembles a winding mode only far from the tip of the cone.

Another hint in this direction has been found in ref.~\cite{David}. The
authors consider type~II theories compactified on twisted circles,
that interpolate between type~II on an ordinary circle and type~0
theories, and in a supergravity analysis they find a tachyon carrying a
non-vanishing momentum in the compact direction, therefore breaking
again the translation invariance and producing a regular lattice.

It is also interesting  to see what  our results suggest for
the condensation of type~0 tachyons.
Type~0 theory is obtained in 
the limit $R\ra 0$, and in this limit the lattice spacing
$a\ra \infty$.
Taking $a\ra\infty$, we are actually
isolating a single spacelike surface. However, 
as discussed in sect.~\ref{unexp}, in this
limit the description in terms of a spacelike surface is not really
adequate: the
$x_i$'s lose their interpretation as coordinates, 
and become boost operators; the
deformed Poincar\'e symmetry on the single $d$-dimensional
spacelike surface at fixed time becomes
equivalent to a full undeformed Poincar\'e symmetry in a $d+1$
dimensional spacetime. However, this is exactly the symmetry of the
ground state of a (non-tachyonic) string theory. 
This result is therefore  consistent with the recent suggestion
that type~0 theories decay to type~II
theories~\cite{CG,GS,Adams,David}. 

\section{Conclusions}

The behavior of strings above the Hagedorn temperature
is, since a long
time, one of the crucial mysteries of string theory. Discussing the
result $F(T)\sim T^2$  that they found, Atick and Witten, back in
1988, commented ``A new version of Heisenberg's principle - some
non-commutativity where it does not usually arise - may be the key to
the thinning of the degrees of freedom that is needed to describe
string theory correctly''\cite{AW}. The results that we have presented
fully vindicate this intuition, and show that the non-commutativity
emerges in a quite subtle way, with a phase transition in which the
Poincar\'e algebra is deformed to a quantum algebra.

The deformations of the Poincar\'e algebra that we have discussed are
also quite interesting structures by themselves. Research in this direction
had somehow got stuck since many years into a (false) problem on the
description of composite systems, see app.~A. 
Once one interprete these deformed
symmetries  properly,
basically as symmetries of the one-particles states only, they offer a
viable and interesting possibility. The results that we have found in 
sect.~2 of this paper show that, at the quantum level, they are the
natural symmetry of system with a discrete time, and that
they are remarkably rich
structures embodying, in a single
conceptually well motivated scheme,
a non-commutative geometry on the spatial coordinates, a generalized
uncertainty principle and a minimal length, deformed dispersion relations
and, in the limit of large deformation parameter, 
a surprising form of time (de)construction. The results of sect.~3
indicate that these ideas can find their place in string theory.

Finally,  in a more general perspective,
the concept of a phase transition where a
symmetry group, rather than being broken to a subgroup, 
is deformed into a non-linear
structure is possibly
interesting in itself and might have broader applications.

\acknowledgments

I am grateful to Maura Brunetti, Maria Alice Gasparini,
Gabriele Veneziano and Konstantin Zarembo
for useful discussions.

\appendix

\section{Composite systems and deformed Poincar\'e symmetry}\label{sectcomp}

In this appendix we clarify a problem on the treatement of
multiparticle systems within the deformed Poincar\'e algebra.
Consider a system made of two particles, with generators
$H_1,{\bf P}_1$ and $H_2,{\bf P}_2$, respectively. 
Since deformed algebras are non-linear structures, if 
$H_1,{\bf P}_1$ and $H_2,{\bf P}_2$ separately satisfy a deformed
algebra, then $H_1+H_2$ and ${\bf P}_1+{\bf P}_2$ do not satisfy the deformed
algebra. 
It has been observed that there is a non-linear
combination of energy and momenta, known as the coproduct,
that satisfies the algebra. For instance, for
the algebra (\ref{PH}) this combination is given by
$E_{12}=\exp\{iaP_2\}E_1+ \exp\{-iaP_1\}E_2$ and
$P_{12}=P_1+P_2$. Using the experience from other types of quantum
groups, in particular deformations of the angular momentum algebra,
there have been attempts to interprete
these expressions as the
energy and momentum of the composite system. However, 
these attempts have immediately failed.
 First of all, $E_{12}$ is not
even real, nor symmetric under the exchange of the two particles, so
that $E_{21}$ defines a different composition law. The
situation does not improve considering any of the other deformed
algebras discussed above. In the algebra (\ref{HP}) now energies add
simply, $E_{12}=E_1+E_2$, but momenta have a non-trivial, and again
physically unacceptable  coproduct,
$P_{12}= \exp\{iaE_2\}P_1+\exp\{-iaE_1\}P_2$.  The algebra
(\ref{euP}) instead has again $E_{12}=E_1+E_2$ and
$P_{12}=\exp\{aE_2\}P_1+ \exp\{-aE_1\}P_2$. This latter
composition law is,
at least, real, but it is still physically meaningless, since it
states that the total momentum of two infinitely separated particles is a
combination (non even symmetric)
of the momenta of one-particle and the
exponential of the energy of the other. The 
point of view that $a$ is very small
and therefore the effects are not observable is
also untenable. Combining a sufficiently large number of particles
with this coproduct rule we would find that even macroscopic objects
obey these unphysical rules for the composition of energy and momenta.

If however we do not lose sight of the physics, the situation is quite
clear. For instance, the algebra (\ref{PH}) is a useful description of
a system of 1+1 phonons. The usefulness comes from the fact that its
Casimir reproduces the wave equation, but certainly phonons do not
satisfy these strange composition laws (while, even in 1+1 dimensions,
the coproduct would give a strange result for two antiparallel
phonons). So this is an explicit counterexample showing that the
deformed Poincar\'e
algebra can be a  useful tool in the
description of a system with a standard
composition law for energy and momenta, 
$E_{\rm tot}=E_1+E_2, P_{\rm tot}=P_1+P_2$.
In other words, $E_{\rm tot}$ and $P_{\rm tot}$ do not have to satisfy
the deformed algebra.

In general, the physical reason why the generators of a composite system 
are expected to satisfy
the same algebra as the constituents is that we can imagine to put 
two elementary constituents into a black box, and there is no way to
distinguish this composite system from an elementary object. 
However, two particles with energy and momenta $E_1,{\bf P}_1$ and
$E_2,{\bf P}_2$ cannot be put in the same finite size black box, 
unless they are
exactly collinear and with the same speed, even more so in 
relativistic quantum
theory, where  particles are only defined as asymptotic
states.
Therefore
the argument which says that the total energy and momentum of the
composite system must
satisfy the same algebra as the constituents
does not go through. The system of two free particles moving in two
different directions is in no way the same as a single localized system.
Of course in the undeformed
case the total energy and momenta still satisfy the Poincar\'e
algebra; however this comes out for free because the algebra is linear. In the
deformed case instead this does not come out automatically, and there
is no argument that requires it, and indeed phonons provide an
explicit counterexample. The deformed Poincar\'e algebras discussed in
sect.~\ref{sect.aP} are
a useful description of discretized systems, and the action of the
generators $P^{\mu}_1, P^{\mu}_2$ on a composite system 
$|1\rangle|2\rangle$ of two particles not exactly collinear
is simply given by
\be
\( e^{-ia_{\mu}P^{\mu}_1}|1\rangle \)
\( e^{-ia_{\mu}P^{\mu}_2}|2\rangle \)
\ee
rather than by $\exp\{-ia_{\mu}P^{\mu}_{12}\}|1\rangle|2\rangle$,
with $P^{\mu}_{12}$ given by the coproduct. As asymptotic states,
$|1\rangle$ and $|2\rangle$ are infinitely separated and the
generators $P^{\mu}_1, P^{\mu}_2$
act separately on their respective single particle states.

There is however a point to be checked. In the
(somewhat idealized) case when  the two particles are {\em exactly}
collinear and with {\em exactly} the same speed, the ``black box argument''
can be applied, and therefore the coproduct must give a meaningful
answer. To understand what is a meaningful answer in this context
observe that, when the dispersion relation is deformed, we cannot have
at the same time the standard relation between momentum and velocity, 
${\bf p}=\g m{\bf v}$, with $\g =(1-v^2)^{-1/2}$, and the 
standard relation between momentum and energy, ${\bf p}=E{\bf v}$,
since one follows from the other upon  use of $E^2={\bf p}^2+m^2$. 
Suppose that ${\bf p}=E{\bf v}$ is modified into
${\bf p}=u(E){\bf v}$, with $u(E)$ a given function. The 
non-trivial coproduct must
simply express the obvious fact that, when $u(E)$ is not linear in
$E$, the momentum of a system with energy $E_{12}=E_1+E_2$ is not just the
sum of the momentum of a system with energy $E_1$ and  
of that of a system with energy $E_2$. Rather,
one has ${\bf P}_{12}=u(E_{12}){\bf v}$.

To verify this, we use for instance the algebra (\ref{elnr1},
\ref{elnr2}), which is
the more interesting for application to string theory.
The coproduct in this case is $E_{12}=E_1+E_2$ and
\be
{\bf P}_{12}=e^{aE_2}{\bf P}_1+ e^{-aE_1}{\bf P}_2 
= ({\bf P}_1 \cosh aE_2 +{\bf P}_2\cosh aE_1) +
({\bf P}_1 \sinh aE_2 -{\bf P}_2\sinh aE_1)\, .\label{P12}
\ee
Thus, we want to find a function $u(E)$ such that, if
we take ${\bf P}_1$ and  ${\bf P}_2$ to be collinear, and with modulus 
$p_1=u(E_1)v$ and $p_2=u(E_2)v$ with {\em the same} $v$, then
\eq{P12} becomes $P_{12}=u(E_1+E_2)v$. It is easy to see that the only
function that does the job (and reduces to $u(E)=E$ if $a\ra 0$)
is $u(E)=\sinh (aE)/a$, i.e.
\be\label{v}
{\bf p}=\frac{\sinh aE}{a}\, {\bf v}\, ,
\ee
since then \eq{P12} becomes
\be
{\bf P}_{12}=\frac{\sinh a(E_1+E_2)}{a}\, {\bf v}\, .
\ee
As a bonus, we have understood that \eq{v} is the correct definition
of the velocity when Poincar\'e symmetry is deformed. Eliminating
$\sinh aE$ with the help of the dispersion relation, we find that this
relation simply means that the standard relation between momentum and
velocity,
${\bf p}=\g m{\bf v}$, with $\g =(1-v^2)^{-1/2}$,
is not deformed~\cite{MM3}. 
The velocity therefore still ranges between zero and
one.
The same procedure can be applied to
any of the other deformations of the Poincar\'e algebra that we have
discussed. For the algebra representing discrete Minkowski time it
suffices to exchange $a\ra ia$, i.e. to define the velocity from
$ap=\sin (aE) v$. Again,  using the dispersion relation, we see that
$p=\g mv$ is not modified.
Then the coproduct for two collinear particles 
with the same speed turns
out to be  real, symmetric under the exchange of the two
particles, and reproduces correctly the non-linearity of the
dispersion relation. In all other
kinematical configurations, the coproduct of the deformed Poincar\'e
algebra has no physical interest.

\end{document}